\documentclass[journal]{IEEEtran}

\usepackage[pdftex]{graphicx}	

\usepackage{latexsym}
\usepackage[fleqn]{amsmath}
\usepackage{amsmath,amssymb}

\usepackage{latexsym}

\renewcommand{\baselinestretch}{1.00}

\usepackage{here}
\usepackage{graphicx}
\usepackage{latexsym}
\usepackage{comment}
\usepackage{multirow}
\usepackage{bm}
\usepackage{cite}
\usepackage[hang,small,bf]{caption}
\usepackage[subrefformat=parens]{subcaption}
\def\lla{\left\langle}
\def\rra{\right\rangle}
\def\lp{\left(}
\def\rp{\right)}

\ifCLASSINFOpdf
\else
\fi
\begin{document}

%
\title{Statistical-mechanical analysis of adaptive filter with clipping saturation-type nonlinearity}
\markboth{
Statistical-mechanical analysis of adaptive filter with clipping saturation-type nonlinearity}{}%
%
%

\author{Seiji~Miyoshi,~\IEEEmembership{Senior~Member,~IEEE}
\thanks{Manuscript received April 19, 202x; revised December 11, 202x.
This work was supported by JSPS KAKENHI 
Grant Number JP20K04494.

S. Miyoshi is with 
the Department of Electrical, Electronic and Information Engineering, 
Faculty of Engineering Science, 
Kansai University, 
3-3-35 Yamate-cho, Suita-shi, Osaka 564-8680, Japan
(e-mail: miyoshi@kansai-u.ac.jp).}
}

\maketitle

\begin{abstract}
In most practical adaptive signal processing systems, e.g.,
active noise control, active vibration control, and 
acoustic echo cancellation,
substantial nonlinearities that cannot be neglected exist.
In this paper, we analyze the behaviors of an adaptive system
in which the output of the adaptive filter 
has the clipping saturation-type nonlinearity
by a statistical-mechanical method. 
We discuss the dynamical and steady-state behaviors 
of the adaptive system
by performing asymptotic analysis, steady-state analysis, and
numerical calculation.
As a result, it is clarified
that the saturation value has a critical point
at which the system's mean-square stability 
and instability switch.
The obtained theory well explains the strange behaviors around 
the critical point observed in the computer simulation.
Finally, the exact value of the critical point is also derived.
\end{abstract}


\section{Introduction}
\IEEEPARstart{A}{daptive} signal processing is 
used in a wide range of areas such as communication systems 
and acoustic systems\cite{Haykin2002, Sayed2003}. 
Active noise control (ANC)
\cite{Nelson1992,Kuo1996,Kuo1999,Kajikawa2012},
active vibration control (AVC)\cite{Fuller1995},
acoustic echo cancellation\cite{Sondhi2006},
and
system identification\cite{Ljung1999}
are examples of specific applications of 
adaptive signal processing.
Adaptive signal processing using linear filters
has been 
theoretically analyzed 
\cite{Haykin2002, Sayed2003}.

The components of most practical adaptive systems, 
such as power amplifiers and 
transducers such as loudspeakers and microphones,
have substantial nonlinearities 
that cannot be neglected\cite{Haykin2002, Sayed2003}.
Such nonlinearities are inevitable, and
it is extremely important to investigate in detail
their effects on the overall performance of adaptive systems.
%
Therefore, 
there have been many studies on adaptive signal processing systems
including nonlinear components
\cite{Widrow1971, Claasen1981, Duttweiler1982, Aref2015, 
Smaoui2011, Bekrani2014, 
Deivasigamani1982, Bekrani2012, Jun1995, Takahashi1992,
Eweda1990, Bershad1990, Costa2001, Costa1999, Costa2002, 
Snyder1995, Costa2017, Costa2008, Tobias2006, Bershad2009, 
Hamidi2004, Stenger2000
}.
In some of these studies,
nonlinearities where an input signal and an error signal
are expressed by their signs ($\pm 1$) or 
three values ($-1, 0, +1$) 
have been investigated 
\cite{Widrow1971,Claasen1981,Duttweiler1982,
Aref2015,Smaoui2011,Bekrani2014,Deivasigamani1982,Bekrani2012,
Jun1995,Takahashi1992,Eweda1990}.
Note, however, 
that such nonlinearities are intended to reduce computational complexity.
Bershad\cite{Bershad1990} analyzed 
the case in which the update by 
the least-mean-square (LMS) algorithm
\cite{Widrow1960}
has $\left(1-e^{-ax}\right)$ saturation-type nonlinearity,
assuming a small step size. 
Costa {\it et al.}\cite{Costa2001} analyzed the case in which
the output of the adaptive filter has an error function (erf)
saturation-type nonlinearity,
assuming a small step size. 
Costa {\it et al.}\cite{Costa1999,Costa2002} analyzed
ANC in which the secondary path
has an erf saturation-type nonlinearity.  
Snyder and Tanaka\cite{Snyder1995} proposed
the replacement of the finite-duration impulse response (FIR) filter
with a neural network
to deal with the primary path nonlinearity in ANC/AVC. 
Costa\cite{Costa2017}
analyzed a hearing aid feedback canceller with 
an erf saturation-type nonlinearity.
Costa {\it et al.}\cite{Costa2008}
analyzed a model in which the output of the adaptive filter
has a dead-zone nonlinearity
caused by a class B amplifier or a nonlinear actuator,
assuming a small step size. 
Tobias and Seara\cite{Tobias2006}
analyzed the behaviors of the modified LMS algorithm
derived from the improved cost function
in the case where
the output of the adaptive filter has
an erf saturation-type nonlinearity.
Bershad\cite{Bershad2009}
analyzed the case where the update by the LMS algorithm
has an erf saturation-type nonlinearity
and extended the analysis to the tracking of
a Markov channel in the context of system identification.
As described so far,
there have been many studies on 
adaptive systems
with an erf saturation-type nonlinearity.
On the other hand,
Hamidi {\it et al.}\cite{Hamidi2004}
reported their analysis, computer simulation, and experimental results
of an ANC model in which the output of the adaptive filter
has the clipping saturation-type nonlinearity.
They proposed a modification of 
the cost function to avoid using a nonlinear region
to improve the adaptive algorithm.
Stenger and Kellermann\cite{Stenger2000}
proposed the use of clipping-type preprocessing
in adaptive echo cancellation
to cancel the effect of nonlinear echo paths.

On the other hand, our group has recently studied 
the application of the statistical-mechanical method \cite{Nishimori2001} 
to the analysis of adaptive signal processing.
In traditional statistical analysis, it is generally 
necessary to use a number of approximations and assumptions to compute 
expectations with respect to the input signal, 
which is a random variable.
On the other hand, in statistical-mechanical analysis,
by considering the large-system limit, 
the universal properties of a system 
consisting of many microscopic variables 
can be simply discussed macroscopically and deterministically 
in terms of a small number of macroscopic variables.
In addition, since the law of large numbers and the central limit theorem 
hold, many calculations required in the analysis become easy to perform.
Statistical-mechanical analysis
is particularly suitable for the analysis of signal processing 
that involves an adaptive filter with a large tap length
as is common in practical acoustic systems.
Note that 
we are not implying that 
statistical-mechanical analysis is superior
to traditional statistical analysis
since the large-system-limit assumption can be a weakness
in some cases. 
Our group \cite{Miyoshi2011EL, Miyoshi2018} 
has also analyzed feed-forward 
ANC updated by the Filtered-X LMS (FXLMS) algorithm 
using the statistical-mechanical method.
However, the analyses dealt with the case where the primary path, 
secondary path, and adaptive filter were all linear.
Our group \cite{Motonaka2021}
has also 
analyzed a model in which both the unknown system
and adaptive filter have the Volterra-type nonlinearity
\cite{Mathews1991}
as an application of the statistical-mechanical method to nonlinear
adaptive signal processing.
Although Volterra filters have nonlinear characteristics, 
it was relatively easy to apply the statistical-mechanical method
used for linear systems to their analysis.
However, the method was only applicable to Volterra filters of a specific order.
Moreover, in the previous study, we did not deal with simple nonlinearities such as 
saturation and dead-zone types, which are often found in 
actual adaptive processing systems.

As described above, 
there have been many studies on 
adaptive signal processing with nonlinear components.
The erf-type saturation function has been well analyzed in previous studies. 
However, the clipping saturation type, i.e., the piecewise linear type,
is an important alternative expression for 
the saturation phenomena
of components constituting adaptive systems
such as power amplifiers and transducers such as loudspeakers and microphones.
As evidence of this, in their paper\cite{Costa2008} 
on the analysis of the dead-zone type 
rather than the saturation type, Costa {\it et al.} stated 
that to facilitate the development of analytical models, 
it is convenient to approximate the piecewise nonlinearity by
a continuous and more mathematically tractable function.
Moreover, they\cite{Costa2008} compared the results of 
the erf-type analysis with the results of computer simulations performed 
with piecewise linearity.
However, there have been few studies on
clipping saturation-type nonlinearity; 
in particular, there have been no analytical studies
to the best of our knowledge.

In light of these developments,
in this paper,
we analyze the behaviors of a system
with an adaptive filter whose output has 
clipping saturation-type nonlinearity, i.e., piecewise linearity.
The main contributions of this paper are as follows:
\begin{itemize}
\item We analyze the dynamical and steady-state 
behaviors of an adaptive system
in which the output of the adaptive filter 
has the clipping saturation-type nonlinearity,
which is an alternative expression for saturation phenomena
other than erf-type saturation,
by the statistical-mechanical method.
\item To this end, we introduce two macroscopic variables 
to represent the macroscopic state of the system
and derive the simultaneous differential equations
that describe system behaviors
in a deterministic and closed form
under the long-filter assumption.
Although the derived equations
cannot be analytically solved,
we discuss the dynamical behaviors and steady state
of the adaptive system
by performing numerical calculation, asymptotic analysis, 
and steady-state analysis.
In the analysis,
we do not assume that the step size is small.
\item It is clarified
that the saturation value $S$ has a critical value $S_C$
at which the system's mean-square stability 
and instability switch.
That is, when $S>S_C$, 
both the mean-square error (MSE) 
and mean-square deviation (MSD)
converge, i.e., the adaptive system is mean-square stable.
On the other hand, when $S<S_C$,
the MSD diverges, 
i.e., the adaptive system is not mean-square stable.
However, even when $S<S_C$, the MSE converges.
The converged value is a quadratic function of $S$
and does not depend on the step size.
Finally, an exact expression for $S_C$ is derived by asymptotic analysis.
\item The proposed theory not only replaces 
the erf-type nonlinearity analyzed in \cite{Costa2001} 
with a clipping saturation-type nonlinearity, 
but also be generalized to large step sizes and 
to the case where the nonlinearity is so strong 
that the mean-square stability of the system does not hold.
\end{itemize}
%

The rest of this paper is organized as follows.
In Sec. \ref{sec:model}, we define the model analyzed in 
this study.
In Sec. \ref{sec:analysis}, we describe the statistical-mechanical
analysis for the model in detail.
In Sec. \ref{sec:results_and_discussion}, we demonstrate
the validity of the obtained theory
by comparing theoretical results with simulation results.
We also clarify that 
there exists a critical value $S_C$ 
for the saturation value $S$ at which 
the properties of the adaptive system switch markedly.
In addition, some results obtained by steady-state, 
asymptotic, and numerical analyses are shown.
Furthermore, we obtain the exact value of $S_C$
by asymptotic analysis.
In Sec. \ref{sec:conclusions} we conclude this paper.

{\it Notation:}
Scalars are denoted by lowercase fonts.
Exceptionally, $Q$, $S$, $S_C$, $N$, and $M$ are also scalars
in accordance with the conventions used in the corresponding literature. 
Column vectors are denoted by bold lowercase 
and matrices by bold uppercase fonts. 
Superscripts ${}^\top$ and ${}^{-1}$ denote transpose and inverse, 
respectively 
while $\langle \cdot \rangle$ stands for expectation.
Finally, if $\bm{x}$ is a vector, then $\|\bm{x}\|_2^2=\bm{x}^{\top}\bm{x}$.

\section{Model} \label{sec:model}
\begin{figure}[htbp]
\centering
\includegraphics[width=0.850\linewidth,keepaspectratio]{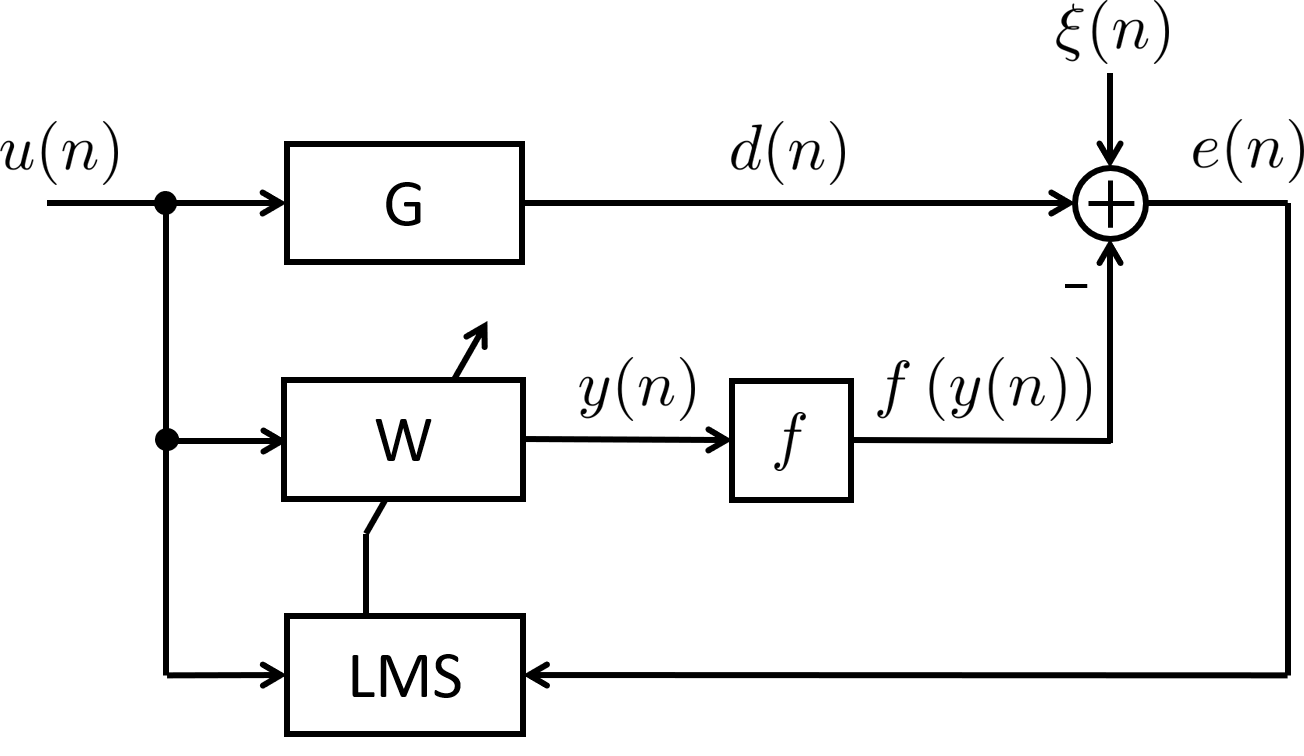}
\caption{Block diagram of the adaptive system.}\label{fig:block}
\end{figure}

Figure \ref{fig:block} shows a block diagram of 
the adaptive system analyzed in this paper.
The impulse response of the unknown system G is an 
$M$-dimensional arbitrary vector
\begin{align}
\bm{g}_0&=[g_1,g_2,\ldots,g_M]^\top,
\end{align}
and is time-invariant.
%
The adaptive filter W is an $N$-tap finite-duration impulse response (FIR) filter.
Its coefficient vector is 
\begin{align}
\bm{w}(n)&=[w_1(n),w_2(n),\ldots,w_N(n)]^\top,
\end{align}
where $n$ denotes the time step.
%
Although the dimension $M$ of $\bm{g}_0$ is generally unknown
in advance, 
we assume that 
the tap length $N$ of the adaptive filter W is set to satisfy
\begin{align}
N\geq M,
\end{align}
because it is straightforward to design 
an adaptive filter W of tap length $N$ with a margin.
Also, let $\bm{g}$ be a vector made into $N$ dimensions 
by adding $N-M$ zeros to $\bm{g}_0$. That is,
\begin{align}
\bm{g}&=[g_1,g_2,\ldots,g_M,g_{M+1},\ldots,g_N]^\top,\\
g_i&=0, \ \ i=M+1,\ldots,N.
\end{align}
Note that while it is assumed in most previous studies 
that the dimensions of $\bm{g}_0$ and $\bm{w}$ are the same
\cite{Costa2001,Costa1999,Costa2002,Costa2008,Tobias2006,Bershad2009},
our model essentially allows an arbitrary $\bm{g}_0$ and 
does not make strict assumptions on its dimension $M$ 
or its elements $\{g_i\}, i=1,\ldots,M$.

We define the parameter $\sigma_g^2$ as
\begin{align}
\sigma_g^2&\triangleq \frac{1}{N}\|\bm{g}_0\|_2^2=\frac{1}{N}\|\bm{g}\|_2^2=
\frac{1}{N}\sum_{i=1}^N g_i^2.
\label{eqn:sigmag2}
\end{align}
As will become clear later, our theory depends on 
the unknown system G only via $\sigma_g^2$.
To demonstrate this, we will show in 
Sec. IV 
that the theory is valid for actual $\bm{g}_0$ 
obtained experimentally.

The input signal $u(n)$
is assumed to be independently drawn from a distribution with
\begin{align}
\left\langle u(n) \right\rangle &=0, \hspace{10mm}
\left\langle u(n)^2 \right\rangle =\sigma^2.
\label{eqn:u_mean_variance}
\end{align}
That is, the input signal is white.
Although the assumption that the input signal is white may seem to 
be a major constraint in this analysis,
the white input model is an important part of practical applications, 
especially in system identification, 
and provides a clear insight into the behavior of the algorithm.
Moreover, the white input case provides a baseline 
for other cases\cite{Costa2001}.
The tap input vector ${\bm u}(n)$ at time step $n$ is 
\begin{align}
\bm{u}(n)&=[u(n),u(n-1),\ldots,u(n-N+1)]^\top.
\end{align}
Note that only the mean and variance of 
the distribution are specified in (\ref{eqn:u_mean_variance}).
No specific distributions, for example, the Gaussian distribution,
are assumed.

Outputs of the unknown system G and the adaptive filter W 
are convolutions of their own coefficients 
and a sequence of input signals.
That is, the outputs $d(n)$ of G
and 
$y(n)$ of W
are respectively
\begin{align}
d(n)&= \bm{g}^\top \bm{u}(n)   = \sum_{i=1}^N g_i    u(n-i+1), \label{eqn:d}\\
y(n)&= \bm{w}(n)^\top \bm{u}(n)= \sum_{i=1}^N w_i(n) u(n-i+1).  \label{eqn:y}
\end{align}

\begin{figure}[htbp]
\centering
\includegraphics[width=0.5\linewidth,keepaspectratio]{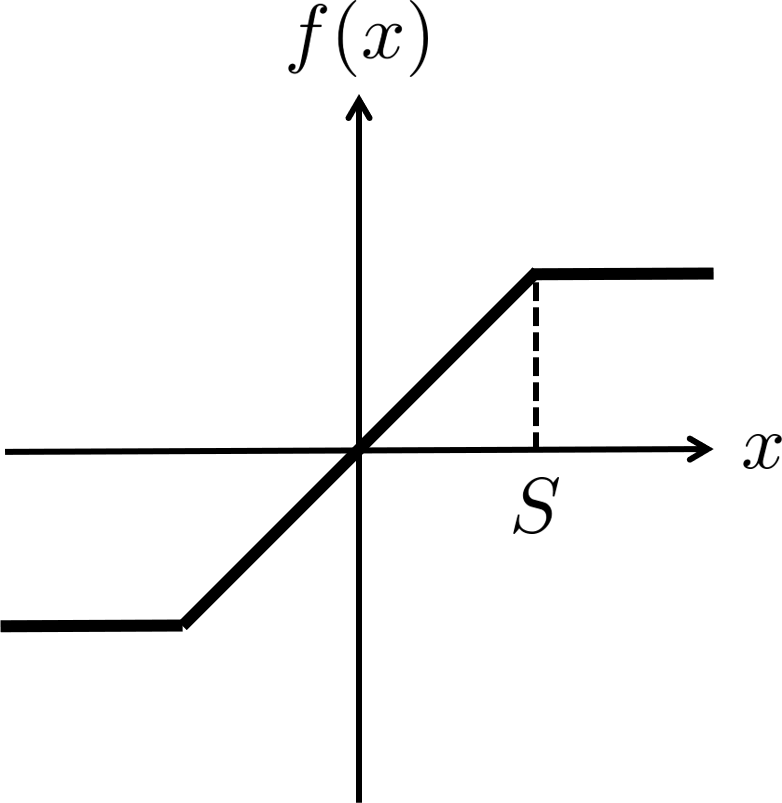}
\caption{Clipping saturation-type nonlinearity.}\label{fig:nonlinear}
\end{figure}

The nonlinearity of the adaptive filter W is modeled by
the function 
$f$ placed after W. 
The function $f$ represents the clipping saturation-type nonlinearity
and is expressed as 
\begin{align}
f(x)&=
	\begin{cases}
		S, 		& 	x>S\\
		-S, 	& 	x < -S\\
		x, 		& 	\mbox{otherwise}
	\end{cases}
	,
\end{align}
where $S$ is a nonnegative real number.
Figure \ref{fig:nonlinear} shows the function $f$.

The error signal $e(n)$ is generated by adding 
an independent background noise $\xi(n)$
to the difference between $d(n)$ and $y(n)$.
That is, 
\begin{align}
e(n)&=d(n)-f(y(n))+\xi(n).	\label{eqn:error}
\end{align}
Here, the mean and variance of $\xi(n)$ are 
zero and $\sigma_\xi^2$, respectively.
Note that only the mean and variance of 
the distribution are specified.
No specific distributions, for example, the Gaussian distribution,
are assumed for the background noise.

The LMS algorithm\cite{Widrow1960}
is used to update the adaptive filter.
That is,
\begin{align}
\bm{w}(n+1)	&=	\bm{w}(n)+\mu e(n)\bm{u}(n),		\label{eqn:LMS}
\end{align}
where $\mu$ is a positive real number and is called the step size.

\section{Analysis} \label{sec:analysis}
In this section, we theoretically analyze the behaviors
of the adaptive system with clipping saturation-type nonlinearity 
by the statistical-mechanical method.
From (\ref{eqn:error}), the MSE is expressed as
\begin{align}
\left\langle e^2\right\rangle 
&= \left\langle \left( d-f(y)+\xi\right)^2\right\rangle\\
&=  \left\langle d^2\right\rangle+\left\langle f(y)^2\right\rangle
-2\left\langle df(y)\right\rangle+\sigma_\xi^2.	\label{eqn:MSE1}
\end{align}
In this section, we omit the time step $n$
unless otherwise stated
to avoid a rather cumbersome notation.
We assume 
$N \to \infty$\footnote{This is called 
{\it the thermodynamic limit} in statistical mechanics.} 
while keeping both $\sigma_g^2$ and 
\begin{align}
\rho^2 &\triangleq N\sigma^2 \label{eqn:rho}
\end{align}
constant, in accordance with the statistical-mechanical
method\cite{Nishimori2001}.

The normalized LMS (NLMS) algorithm\cite{Haykin2002,Sayed2003} 
is a practically important variant of 
the LMS algorithm, whose update rule is
\begin{align}
\bm{w}(n+1)	&=	\bm{w}(n)+\frac{\tilde{\mu}}{\|\bm{u}(n)\|_2^2} e(n)\bm{u}(n),		
\label{eqn:NLMS}
\end{align}
where $\tilde{\mu}$ is the step size.
Note that since $\|\bm{u}(n)\|^2_2=N\sigma^2=\rho^2$, 
the analysis in this paper is equivalent to the analysis of 
the NLMS algorithm with $\tilde{\mu}=\rho^2\mu$ as the step size
for a stationary input signal $u(n)$.

Then, from the central limit theorem, 
both $d$ and $y$ are stochastic variables that obey
the Gaussian distribution.
Their means are zero, and  their variance--covariance matrix is
\begin{align}
	\bm{\Sigma}&\triangleq 
	\lla \left[ d-\lla d \rra, y-\lla y \rra\right]^\top 
	\left[ d-\lla d \rra, y-\lla y \rra\right] \rra \\
	&=
	\rho^2
	\begin{pmatrix}
		\sigma_g^2 	& r		\\
		r 				& Q
	\end{pmatrix}.
	\label{eqn:cov}
\end{align}
Here, $Q$ and $r$ are macroscopic variables that are respectively defined as
\begin{align}
Q &\triangleq \frac{1}{N}\bm{w}^\top\bm{w}, \label{eqn:Qdef} \\
r &\triangleq 	\frac{1}{N}\bm{g}^\top\bm{w}.	\label{eqn:rdef}
\end{align}
The derivation of the means and variance--covariance matrix
is given in detail in Appendix \ref{sec:app_mean_cov}.

We obtain three sample means in (\ref{eqn:MSE1}) as follows
by carrying out the Gaussian integration for 
$d$ and $y$: 
\begin{align}
\left\langle d^2\right\rangle 
&=\rho^2\sigma_g^2,	\label{eqn:d2}	\\
\left\langle f(y)^2\right\rangle
&=S^2+\left(\rho^2 Q-S^2\right)\mbox{erf}
\left(\frac{S}{\sqrt{2\rho^2 Q}}\right)		\nonumber \\
&\hspace{4mm} 
-S\sqrt{\frac{2\rho^2 Q}{\pi}}\exp\left(-\frac{S^2}{2\rho^2 Q}\right),	
\label{eqn:fy2} \\
\left\langle df(y)\right\rangle
&=\rho^2 r\ \mbox{erf}\left(\frac{S}{\sqrt{2\rho^2 Q}}\right),	\label{eqn:dfy}
\end{align}
where $\mbox{erf}(\cdot)$ is an error function defined as
\begin{align}
\mbox{erf}(x)
&\triangleq \frac{2}{\sqrt{\pi}}\int_0^x \exp\left(-\tau^2\right)\mathrm{d}\tau.
\end{align}
%
Equation (\ref{eqn:d2}) is easily derived from (\ref{eqn:cov}).
Equations (\ref{eqn:fy2}) and (\ref{eqn:dfy}) 
are derived in detail in Appendices \ref{sec:appf(y)^2} and \ref{sec:appdf(y)},
respectively.

From (\ref{eqn:MSE1}), (\ref{eqn:d2}), (\ref{eqn:fy2}), and (\ref{eqn:dfy}),
we obtain the MSE as
\begin{align}
\left\langle e^2\right\rangle
&=
\rho^2\sigma_g^2+S^2
+\left(\rho^2 Q-2\rho^2 r-S^2\right)
\mbox{erf}\left(\frac{S}{\sqrt{2\rho^2 Q}} \right)\nonumber \\
&\hspace{4mm} 
- S\sqrt{\frac{2\rho^2 Q}{\pi}}\exp\left(-\frac{S^2}{2\rho^2 Q}\right)
+\sigma_\xi^2.	\label{eqn:MSE2}
\end{align}
%
This formula shows that 
the MSE is a function of
the macroscopic variables $Q$ and $r$.
Therefore, 
we derive differential equations that describe
the dynamical behaviors of these variables in the following.
%
Multiplying 
both sides of (\ref{eqn:LMS}) on the left by 
$\bm{g}^\top$
and using (\ref{eqn:rdef}), we obtain
\begin{align}
Nr(n+1) &= Nr(n) + \mu e(n)d(n). 
\label{eqn:Nr0}
\end{align}

We introduce 
time $t$ defined by
\begin{align}
t &\triangleq n/N,
\end{align}
and use it to
represent the adaptive process.
Then, $t$ becomes a continuous variable since the limit
$N\rightarrow \infty$ is considered.
These calculations are in line with 
the statistical-mechanical analysis of online learning\cite{Engel2001}.

If the adaptive filter is updated 
$N\mathrm{d}t$ times in an infinitely small time $\mathrm{d}t$, 
we can obtain $N\mathrm{d}t$ equations as
\begin{align}
Nr(n+1) &= Nr(n)    + \mu e(n)d(n),       \label{eqn:Nr_n+1} \\  
Nr(n+2) &= Nr(n+1) + \mu e(n+1)d(n+1), \label{eqn:Nr_n+2} \\  
\vdots \ \ \ \ \  &\hspace{1.5mm}\vdots \ \ \ \ \ \ \ \vdots \nonumber \\
Nr(n+N\mathrm{d}t) 
    &= Nr(n+N\mathrm{d}t-1) \nonumber \\
    &\hspace{5mm} + \mu e(n+N\mathrm{d}t-1)d(n+N\mathrm{d}t-1). \label{eqn:Nr_n+Ndt}
\end{align}
Summing all these equations,
we obtain
\begin{align}
Nr(n+N\mathrm{d}t) &= Nr(n)+\mu \sum_{n'=n}^{n+N\mathrm{d}t-1}e(n')d(n').
\label{eqn:N(r+Ndt)}
\end{align}
Therefore, we obtain
\begin{align}
N(r+\mathrm{d}r) &= Nr + N\mathrm{d}t \mu \lla ed\rra. \label{eqn:Nr+dr}
\end{align}
Here, 
from the law of large numbers,
we have represented the effect of the probabilistic variables by
their means since the updates are executed 
$N\mathrm{d}t$ times,
that is, many times, to change $r$ by $\mathrm{d}r$.
This property 
is called {\it self-averaging} in statistical mechanics\cite{Nishimori2001}.
From (\ref{eqn:error}) and (\ref{eqn:Nr+dr}),
we obtain a differential equation that describes
the dynamical behavior of $r$ in a deterministic 
form as follows: 
\begin{align}
\frac{\mathrm{d}r}{\mathrm{d}t}
&=\mu \Bigl(\left\langle d^2\right\rangle-\left\langle df(y)\right\rangle\Bigr).
\label{eqn:drdt}
\end{align}

Next, squaring both sides of (\ref{eqn:LMS})
and proceeding 
in the same manner as for the derivation of 
the above differential equation for $r$,
we can derive a differential equation for $Q$,
which is given by
\begin{align}
\frac{\mathrm{d}Q}{\mathrm{d}t}
&=\mu^2 \rho^2 \Bigl(\left\langle d^2\right\rangle-2\left\langle df(y)\right\rangle
+\left\langle f(y)^2\right\rangle+\sigma_\xi^2\Bigr)\nonumber \\
&\hspace{4mm} 
+2\mu\Bigl(\left\langle dy\right\rangle-\left\langle yf(y)\right\rangle\Bigr).
\label{eqn:dQdt}
\end{align}
Here, we used the fact that in the limit of $N\rightarrow \infty$, 
$\mathbf{u}^\top \mathbf{u}=\|\mathbf{u}\|_2^2=\sum_{i=1}^N u(n-i+1)^2$
is no longer a random variable but a constant $\rho^2=N\sigma^2$ 
from the law of large numbers, 
so this can be taken out of the expectation operation $\langle \cdot \rangle$.
This is precisely one of the strengths of 
the statistical-mechanical method, which assumes $N\rightarrow \infty$.


Equations (\ref{eqn:drdt}) and (\ref{eqn:dQdt}) include five sample means.
However, because three of the five means are already given in 
(\ref{eqn:d2})--(\ref{eqn:dfy}),
we obtain the two remaining means as follows by 
carrying out the Gaussian integration for $d$ and $y$:
\begin{align}
\left\langle dy\right\rangle
&=\rho^2 r,	\label{eqn:dy}	\\
\left\langle yf(y)\right\rangle
&=\rho^2 Q \ \mbox{erf}\left(\frac{S}{\sqrt{2\rho^2 Q}}\right).		\label{eqn:yfy}	
\end{align}
Equation (\ref{eqn:dy}) is easily derived from (\ref{eqn:cov}).
Equation (\ref{eqn:yfy}) is derived in detail in Appendix \ref{sec:appyf(y)}.

Substituting (\ref{eqn:d2})--(\ref{eqn:dfy}), (\ref{eqn:dy}), and (\ref{eqn:yfy})
into (\ref{eqn:drdt}) and (\ref{eqn:dQdt}),
we obtain the concrete formulas of the simultaneous differential equations as follows:
%
\begin{align}
\frac{\mathrm{d}r}{\mathrm{d}t}
&=\mu \rho^2 
\left(\sigma_g^2-r\ \mbox{erf}\left(\frac{S}{\sqrt{2\rho^2 Q}} \right)\right),
\label{eqn:drdt2}\\
\frac{\mathrm{d}Q}{\mathrm{d}t}
&=\mu \rho^2 \Bigl(\mu \left(\rho^2 Q-2\rho^2 r-S^2\right)-2Q\Bigr)
\mbox{erf}\left(\frac{S}{\sqrt{2\rho^2 Q}} \right)\nonumber \\
&\hspace{4mm} 
- \mu^2 \rho^2 S \sqrt{\frac{2\rho^2 Q}{\pi}}\exp\left(-\frac{S^2}{2\rho^2 Q}\right)
\nonumber \\
&\hspace{4mm} 
+\mu \rho^2 \left(\mu\left(\rho^2 \sigma_g^2 + S^2+\sigma_\xi^2\right)+2r\right).
\label{eqn:dQdt2}
\end{align}

We numerically solve the derived simultaneous differential equations, 
since they cannot be analytically solved.
Substituting the obtained numerical solution into 
(\ref{eqn:MSE2}), we obtain the MSE learning curves. 

From (\ref{eqn:sigmag2}), (\ref{eqn:Qdef}), and (\ref{eqn:rdef}),
we can also obtain 
the MSD, 
or misalignment,
as a function of the macroscopic variables $Q$ and $r$ as follows:
\begin{align}
\mbox{MSD}
&= \|\bm{g}-\bm{w}\|_2^2 \\
&= \|\bm{g}\|_2^2-2\bm{g}^\top \bm{w}+\|\bm{w}\|_2^2 \\
&= N(\sigma_g^2-2r+Q).	\label{eqn:MSD}
\end{align}
Equation (\ref{eqn:MSD}) shows that
the MSD is proportional to the tap length $N$ in 
the model setting in this paper;
therefore, we normalize the MSD by $N$ and call it
the normalized MSD.

\section{Results and Discussion} \label{sec:results_and_discussion}
\subsection{Learning curves}
\begin{figure}[htbp]
\centering
\includegraphics[width=0.950\linewidth,keepaspectratio]{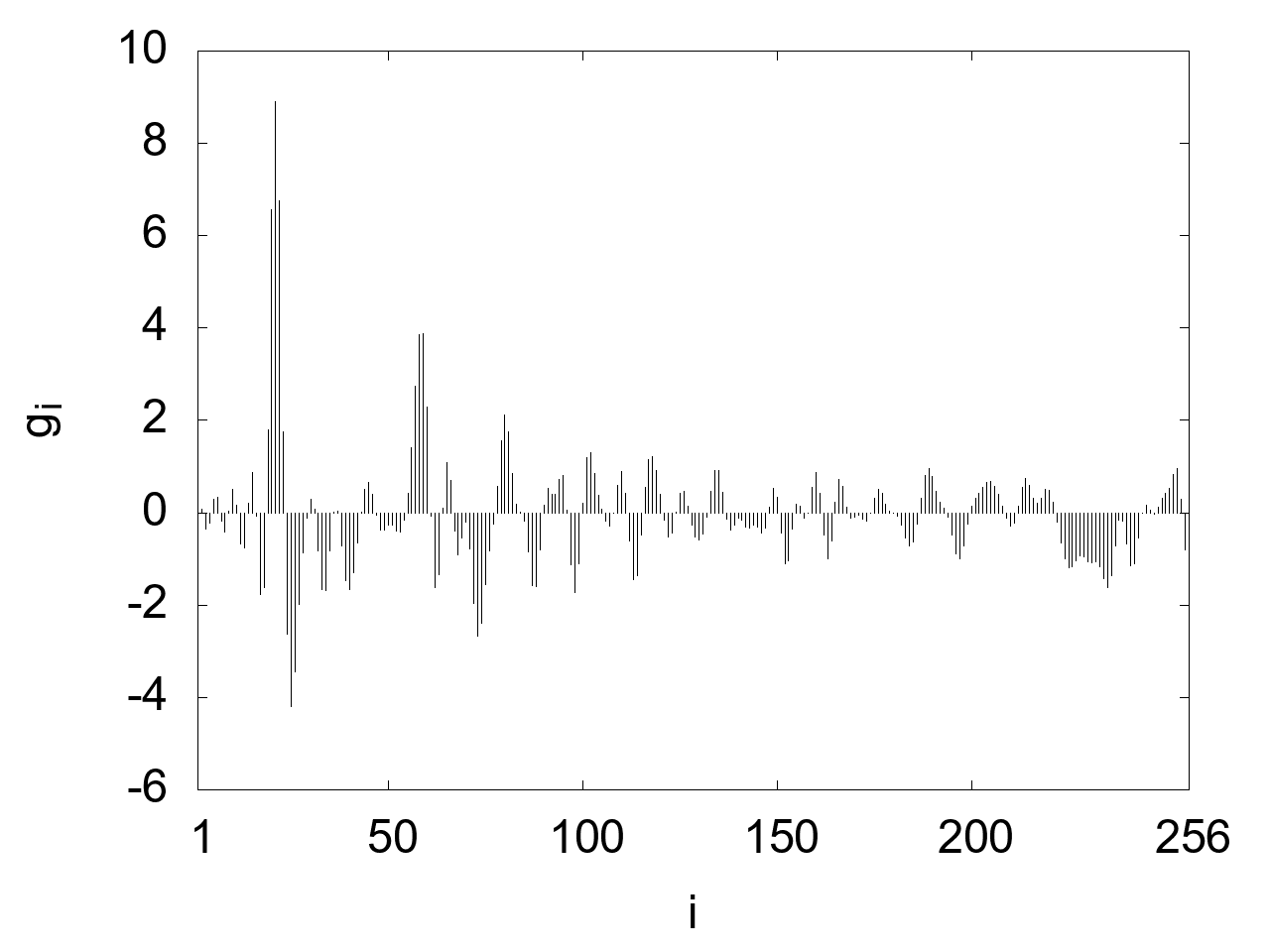}
\caption{Impulse response $\bm{g}_0$ of the unknown system G 
used in all computer simulations in this paper.}\label{fig:realunknown}
\end{figure}

We first investigate the validity of the theory 
by comparing theoretical results with simulation results
with regard to the dynamical behaviors of the MSE and normalized MSD, 
that is, the learning curves.
Figures \ref{fig:MSEVXI0,1} and \ref{fig:MSDVXI0,1} show 
the learning curves obtained 
using the theory derived in the previous section,
along with the corresponding simulation results.
In these figures, the curves represent theoretical results
and the polygonal lines
represent simulation results.
In the theoretical calculation and simulations throughout this section, 
$\rho^2=\sigma_g^2=1$. 
In the theoretical calculation,
the results are obtained by substituting $r$ and $Q$,
which are respectively obtained by solving (\ref{eqn:drdt2}) and (\ref{eqn:dQdt2}),
into (\ref{eqn:MSE2}) and (\ref{eqn:MSD}).
Here, 
(\ref{eqn:drdt2}) and (\ref{eqn:dQdt2}) are numerically solved by
the Runge--Kutta method.
In the computer simulations, 
the number of taps of 
the adaptive filter W is $N=400$ and 
ensemble means for 1000 trials are plotted.
The impulse response $\bm{g}_0$ of the unknown system G 
in all computer simulations in this paper 
is obtained experimentally by measurement\cite{Miyoshi2015ICASSP}
and is shown in Fig. \ref{fig:realunknown}. 
Its dimension $M$ is 256.
Note that $\bm{g}_0$ has been normalized 
to satisfy (\ref{eqn:sigmag2}).
All initial values $w_i(0),\ i=1,\ldots,N$ of the coefficients are 
set to zero in the simulation,
and the initial condition  
$r(0)=Q(0)=0$
is used in the theoretical calculation.
Figures \ref{fig:MSEVXI0,1} and \ref{fig:MSDVXI0,1} show
that the theory derived in this paper
predicts the simulation results well in
terms of average values.

\begin{figure}[htbp]
    \centering
  \begin{minipage}[b]{0.78\linewidth}
    \centering
	\includegraphics[width=1.00\linewidth,keepaspectratio]{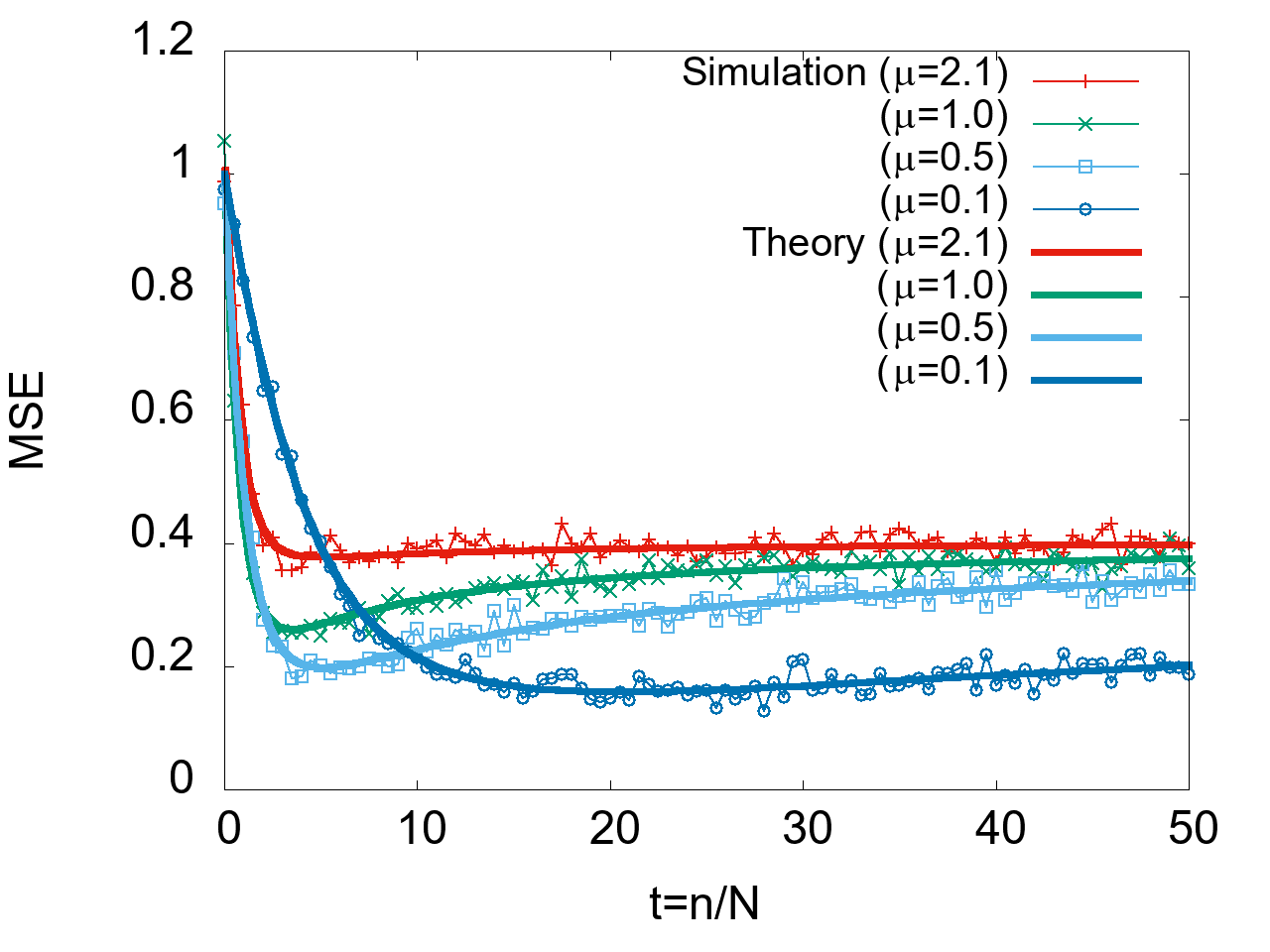}
    \subcaption{$S=1, \sigma_\xi^2=0$}\label{fig:MSES1VXI0RL1}
    \centering
    \includegraphics[width=1.00\linewidth,keepaspectratio]{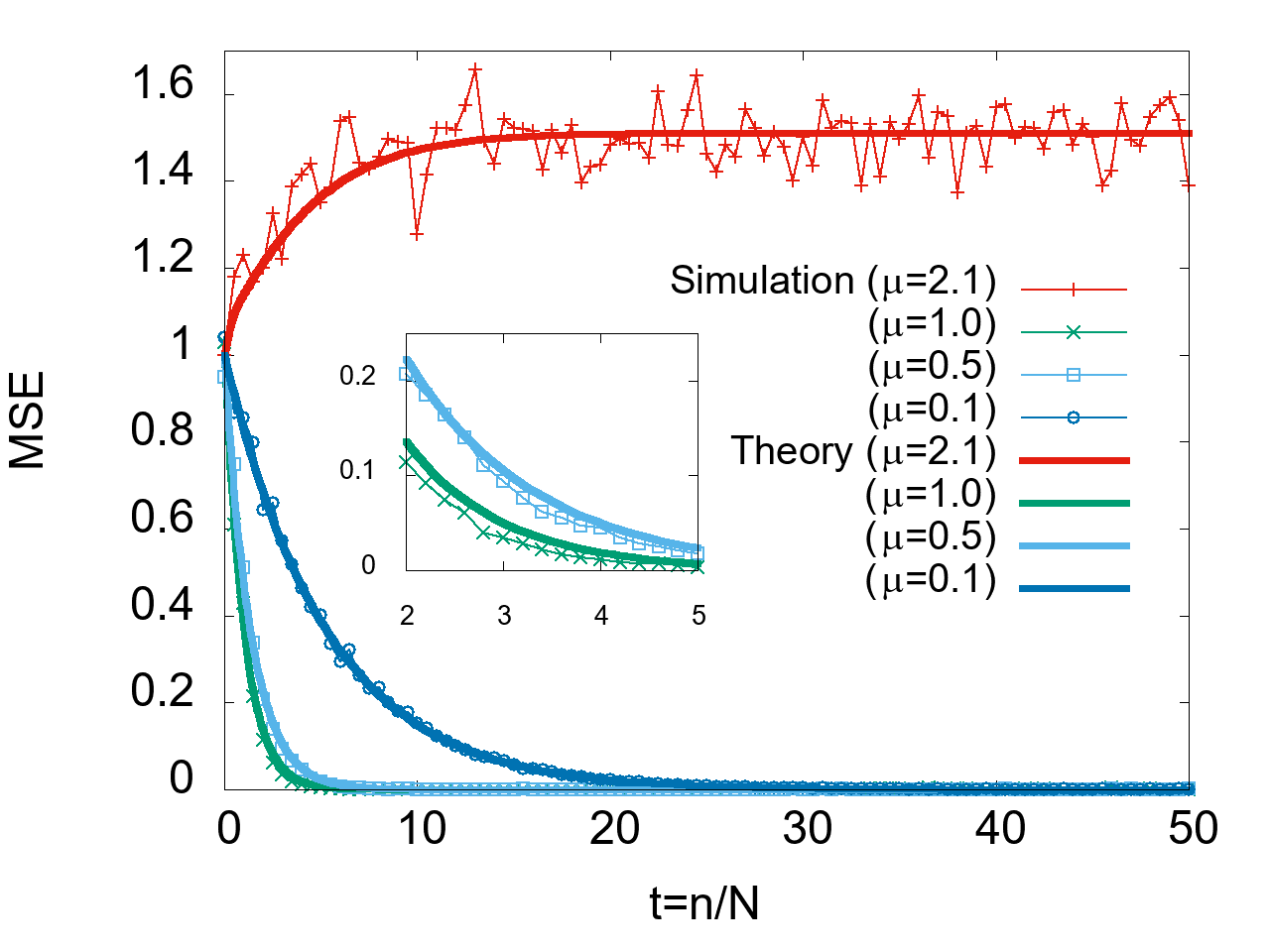}
    \subcaption{$S=3, \sigma_\xi^2=0$}\label{fig:MSES3VXI0RL1}
  \end{minipage}
  \begin{minipage}[b]{0.78\linewidth}
    \centering
	\includegraphics[width=1.00\linewidth,keepaspectratio]{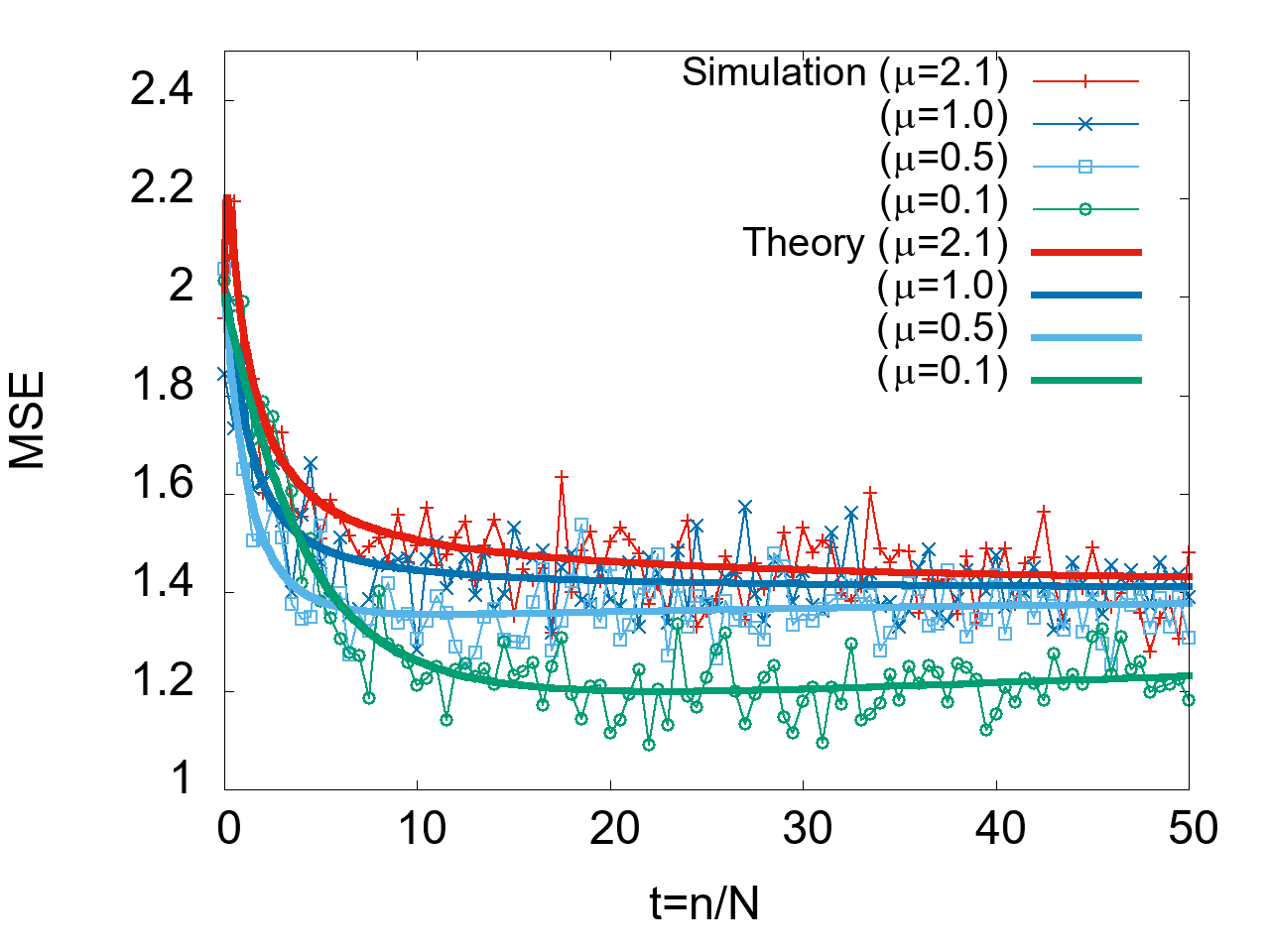}
    \subcaption{$S=1, \sigma_\xi^2=1$}\label{fig:MSES1VXI1RL1}
    \centering
	\includegraphics[width=1.00\linewidth,keepaspectratio]{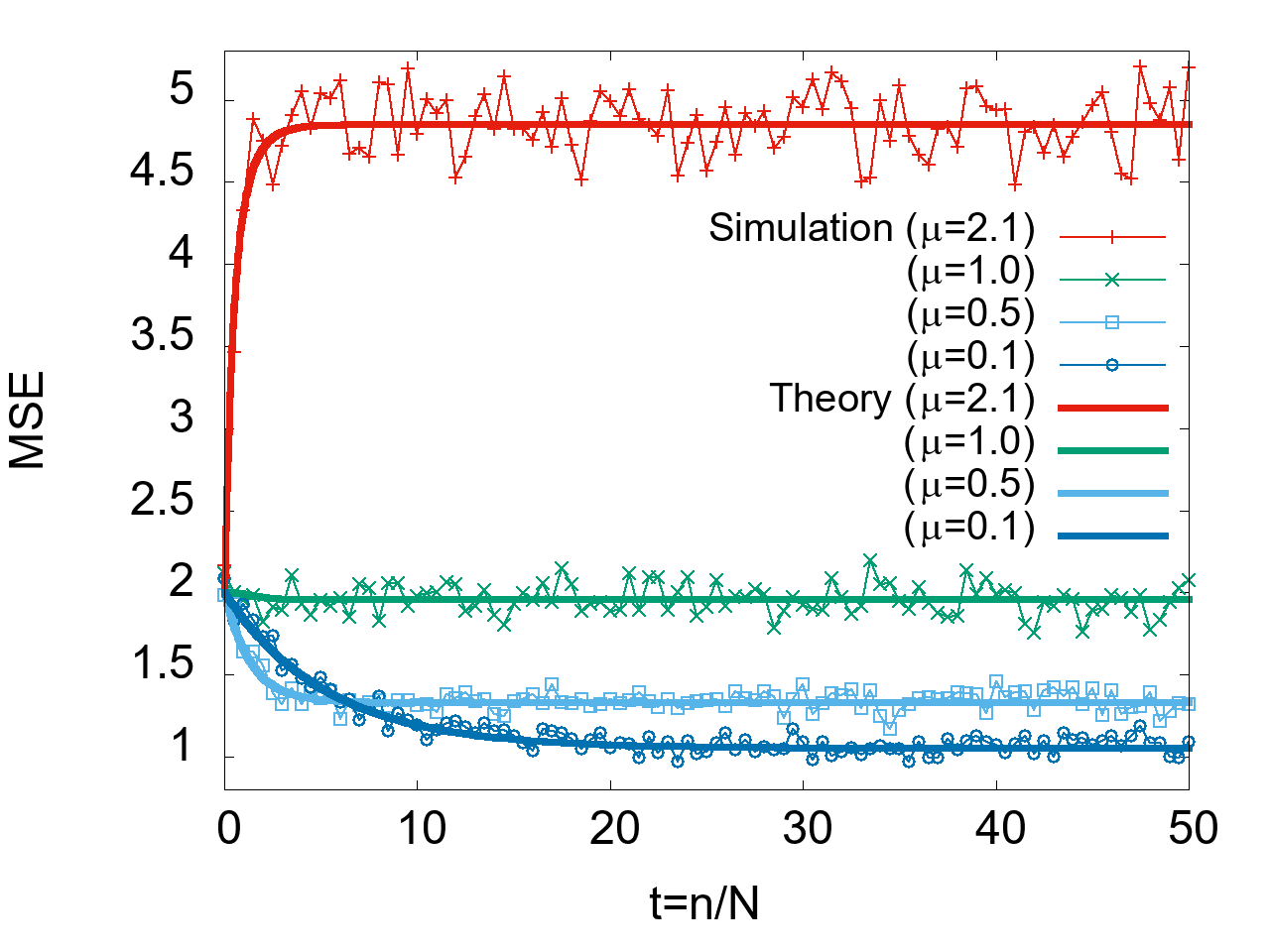}
    \subcaption{$S=3, \sigma_\xi^2=1$}\label{fig:MSES3VXI1RL1}
  \end{minipage}
%
%
  \caption{MSE learning curves.}
  \label{fig:MSEVXI0,1}
\end{figure}

\begin{figure}[htbp]
    \centering
  \begin{minipage}[b]{0.78\linewidth}
    \centering
	\includegraphics[width=1.00\linewidth,keepaspectratio]{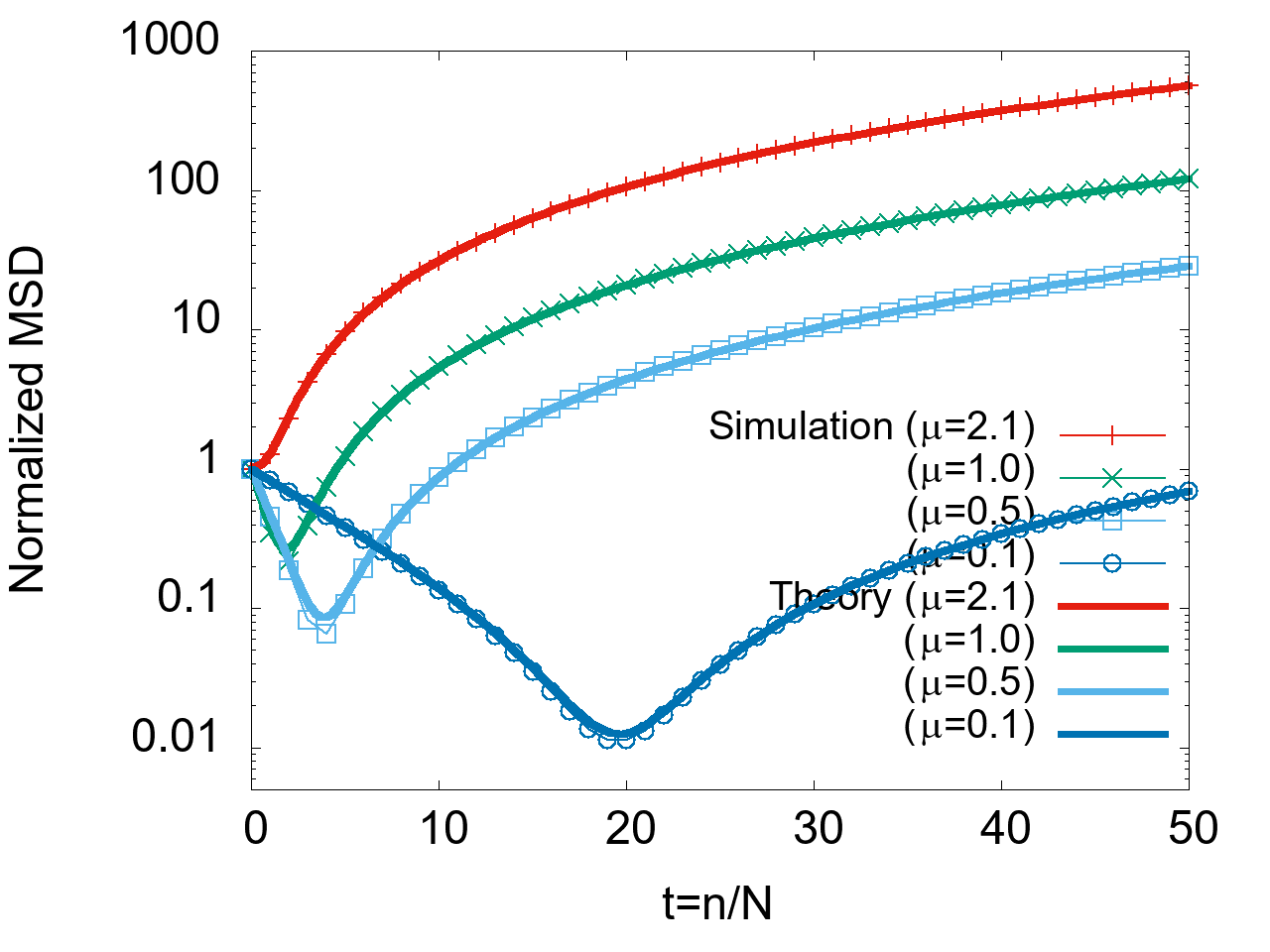}
    \subcaption{$S=1, \sigma_\xi^2=0$}\label{fig:MSDS1VXI0RL1}
    \centering
    \includegraphics[width=1.00\linewidth,keepaspectratio]{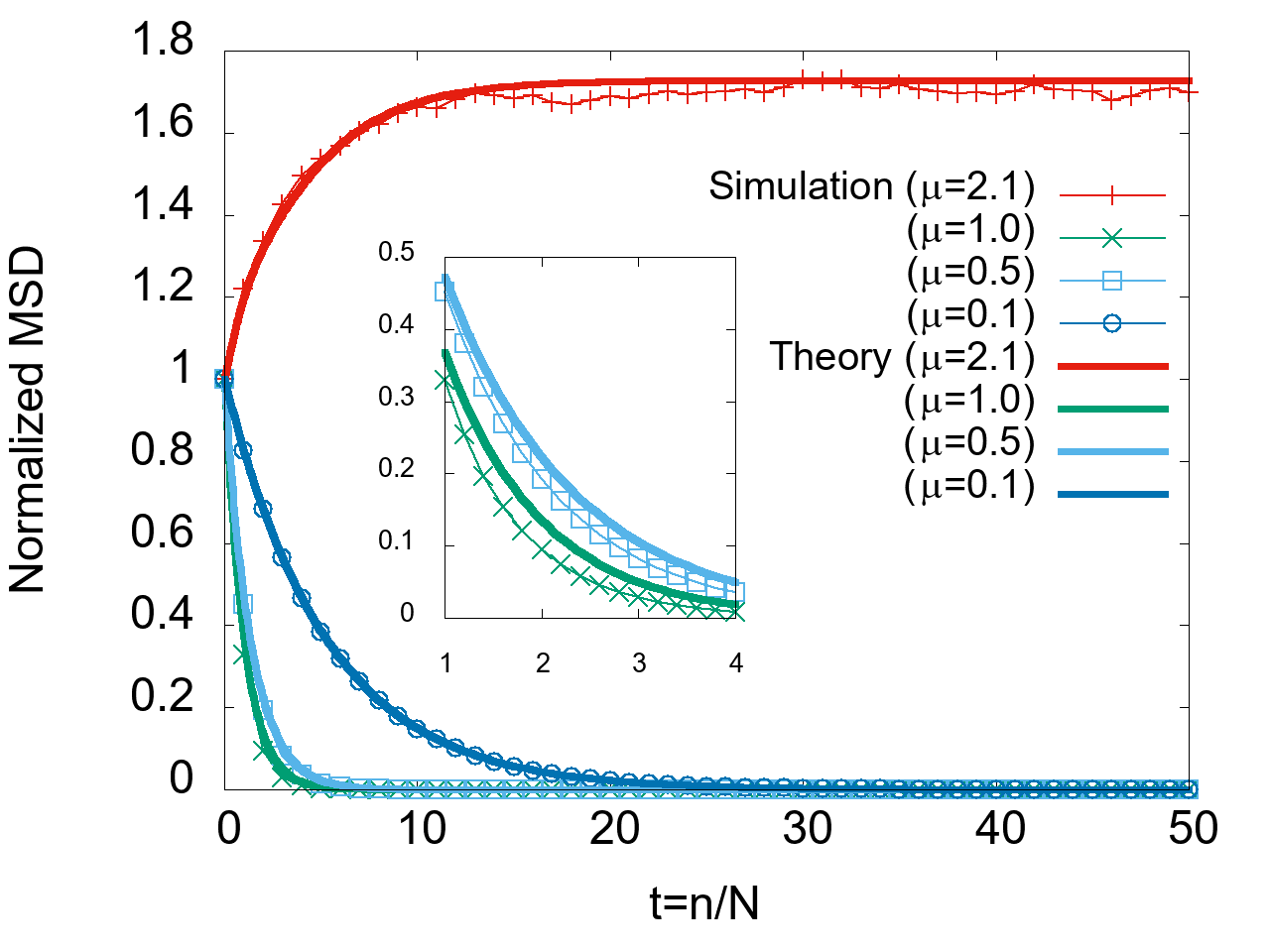}
    \subcaption{$S=3, \sigma_\xi^2=0$}\label{fig:MSDS3VXI0RL1}
  \end{minipage}
  \begin{minipage}[b]{0.78\linewidth}
    \centering
	\includegraphics[width=1.00\linewidth,keepaspectratio]{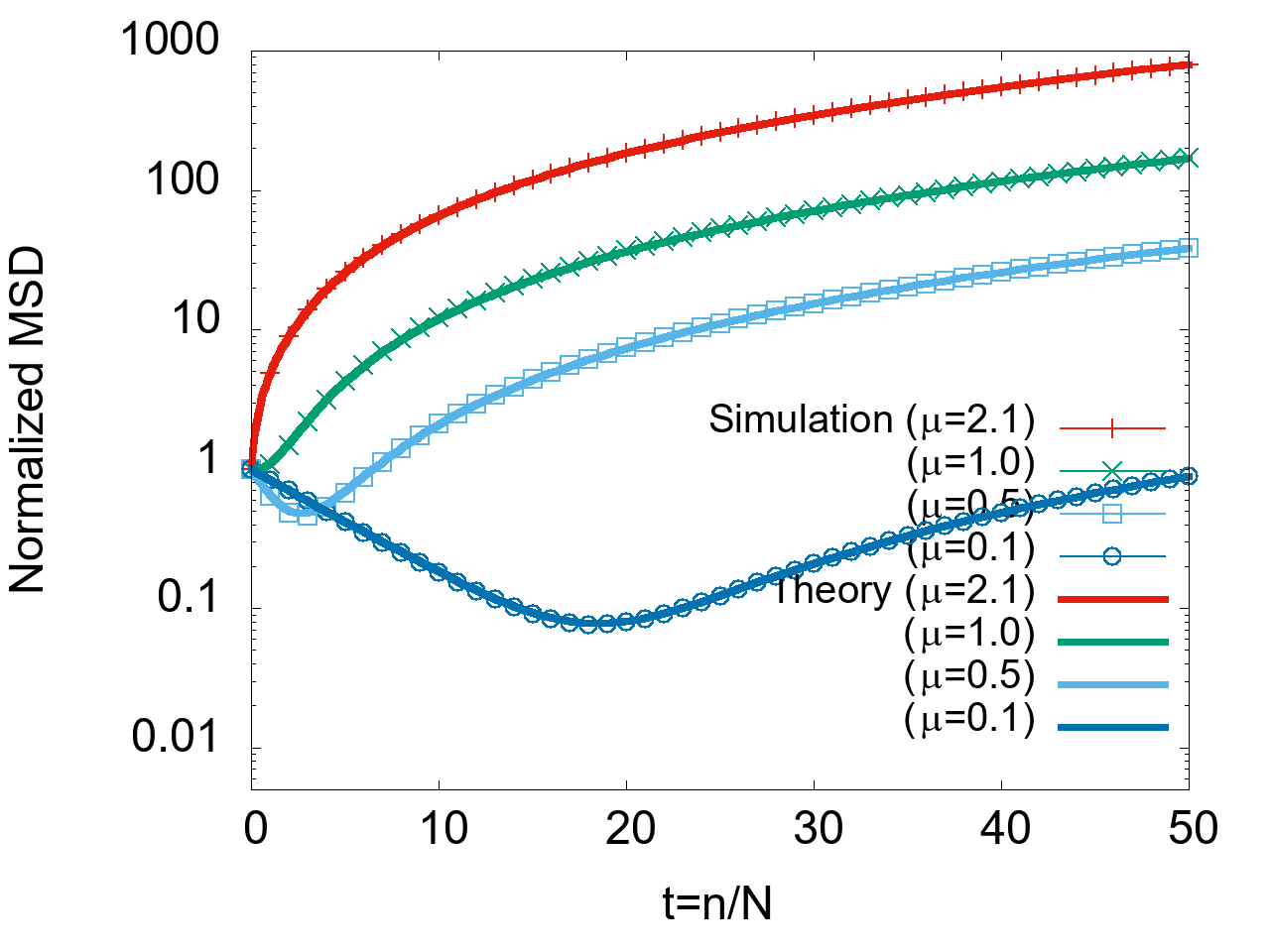}
    \subcaption{$S=1, \sigma_\xi^2=1$}\label{fig:MSDS1VXI1RL1}
    \centering
    \includegraphics[width=1.00\linewidth,keepaspectratio]{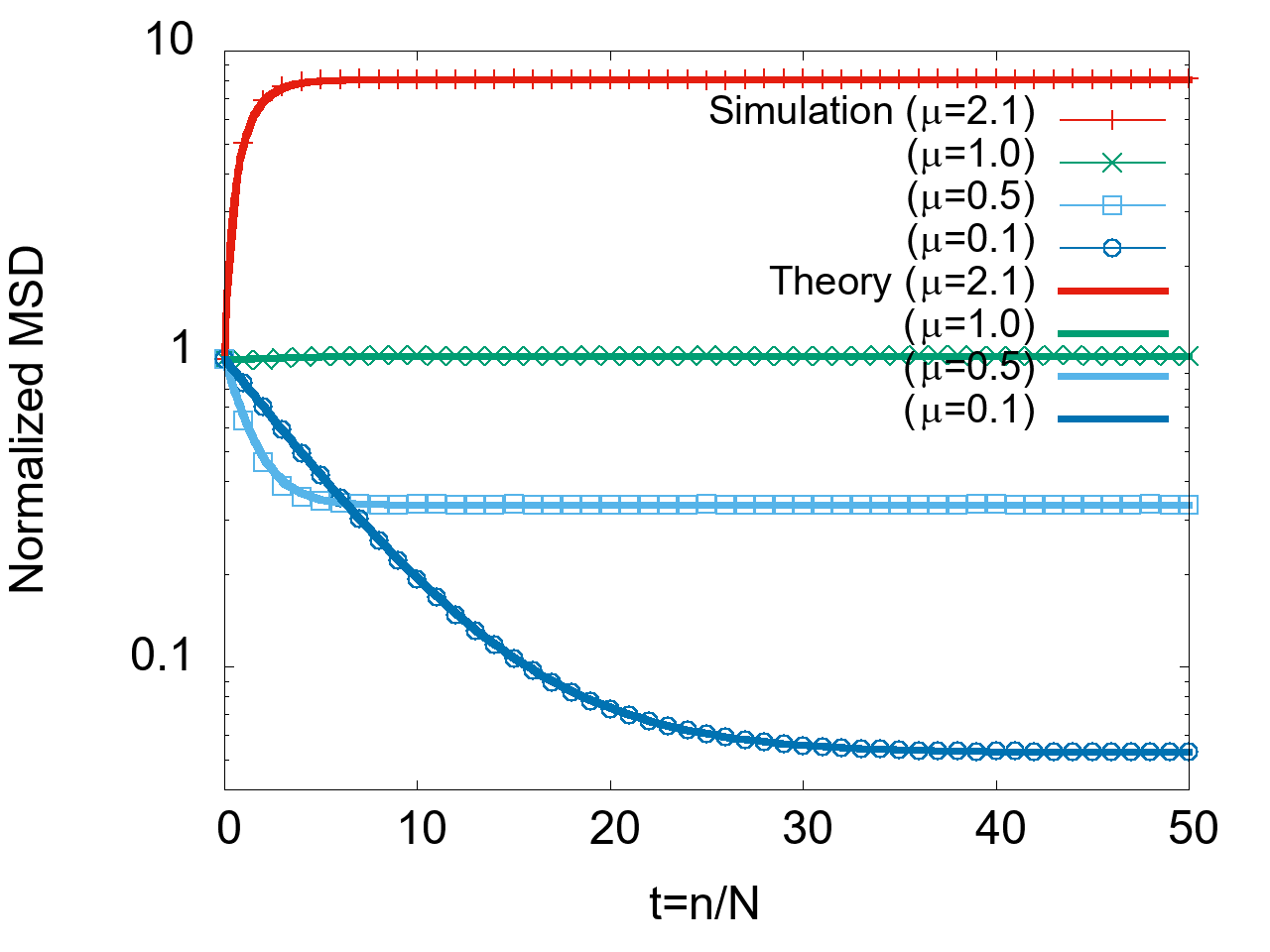}
    \subcaption{$S=3, \sigma_\xi^2=1$}\label{fig:MSDS3VXI1RL1}
  \end{minipage}
%
%
  \caption{Normalized MSD learning curves.}
  \label{fig:MSDVXI0,1}
\end{figure}

\begin{figure}[htbp]
    \centering
  \begin{minipage}[b]{0.78\linewidth}
    \centering
	\includegraphics[width=1.00\linewidth,keepaspectratio]{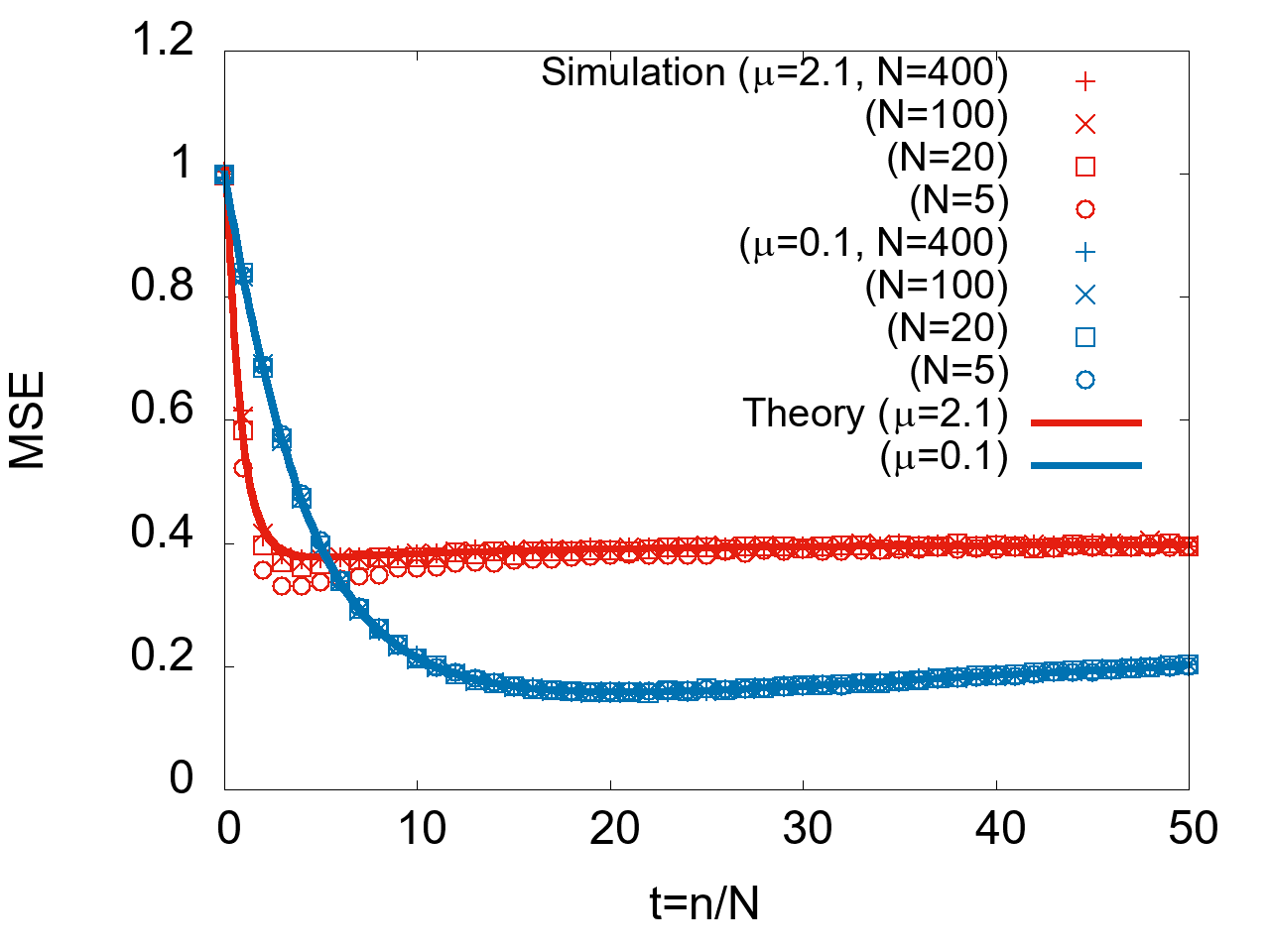}
    \subcaption{$S=1, \sigma_\xi^2=0$}\label{fig:MSES1VXI0FSE}
    \centering
    \includegraphics[width=1.00\linewidth,keepaspectratio]{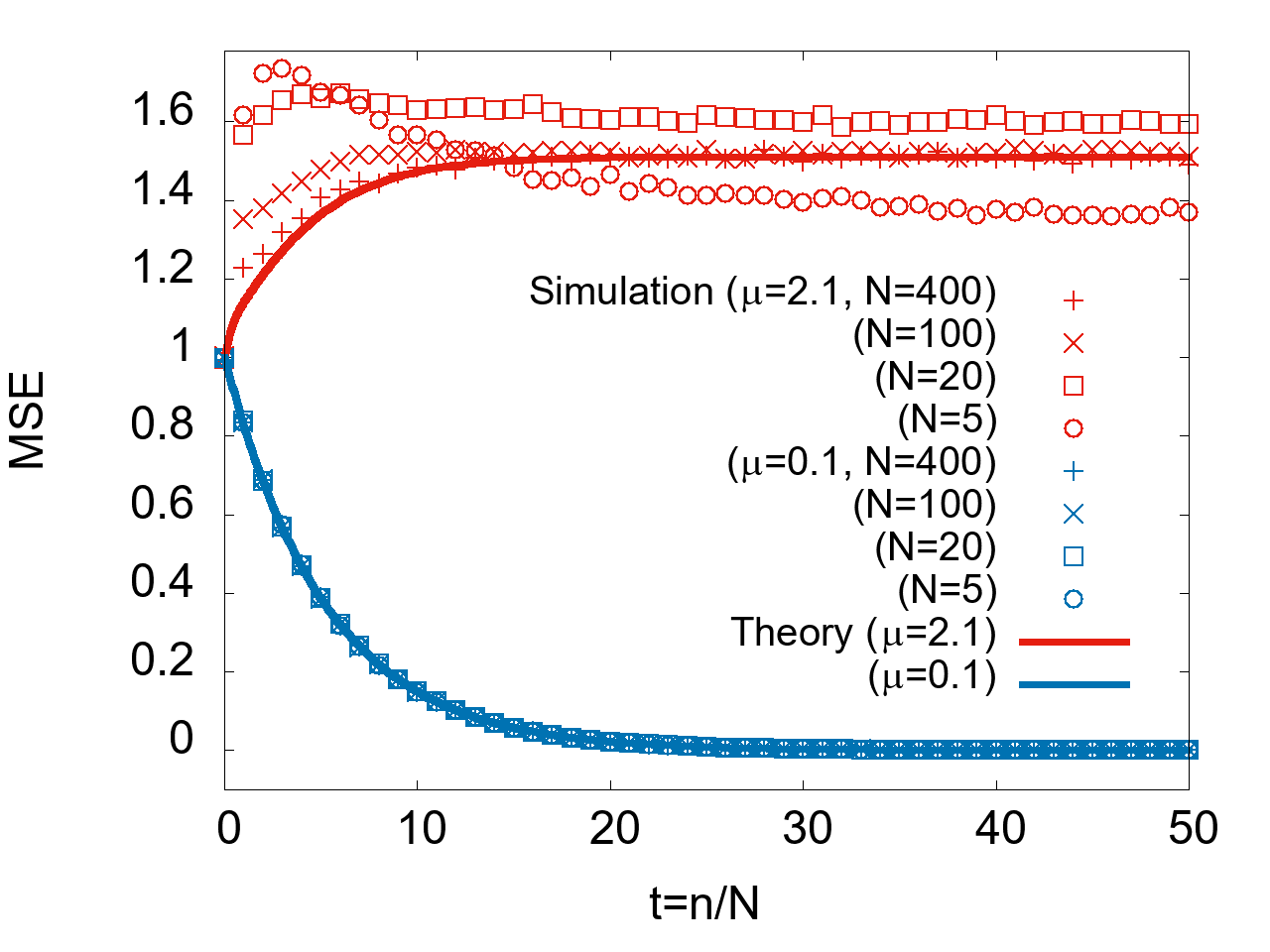}
    \subcaption{$S=3, \sigma_\xi^2=0$}\label{fig:MSES3VXI0FSE}
  \end{minipage}
  \begin{minipage}[b]{0.78\linewidth}
    \centering
	\includegraphics[width=1.00\linewidth,keepaspectratio]{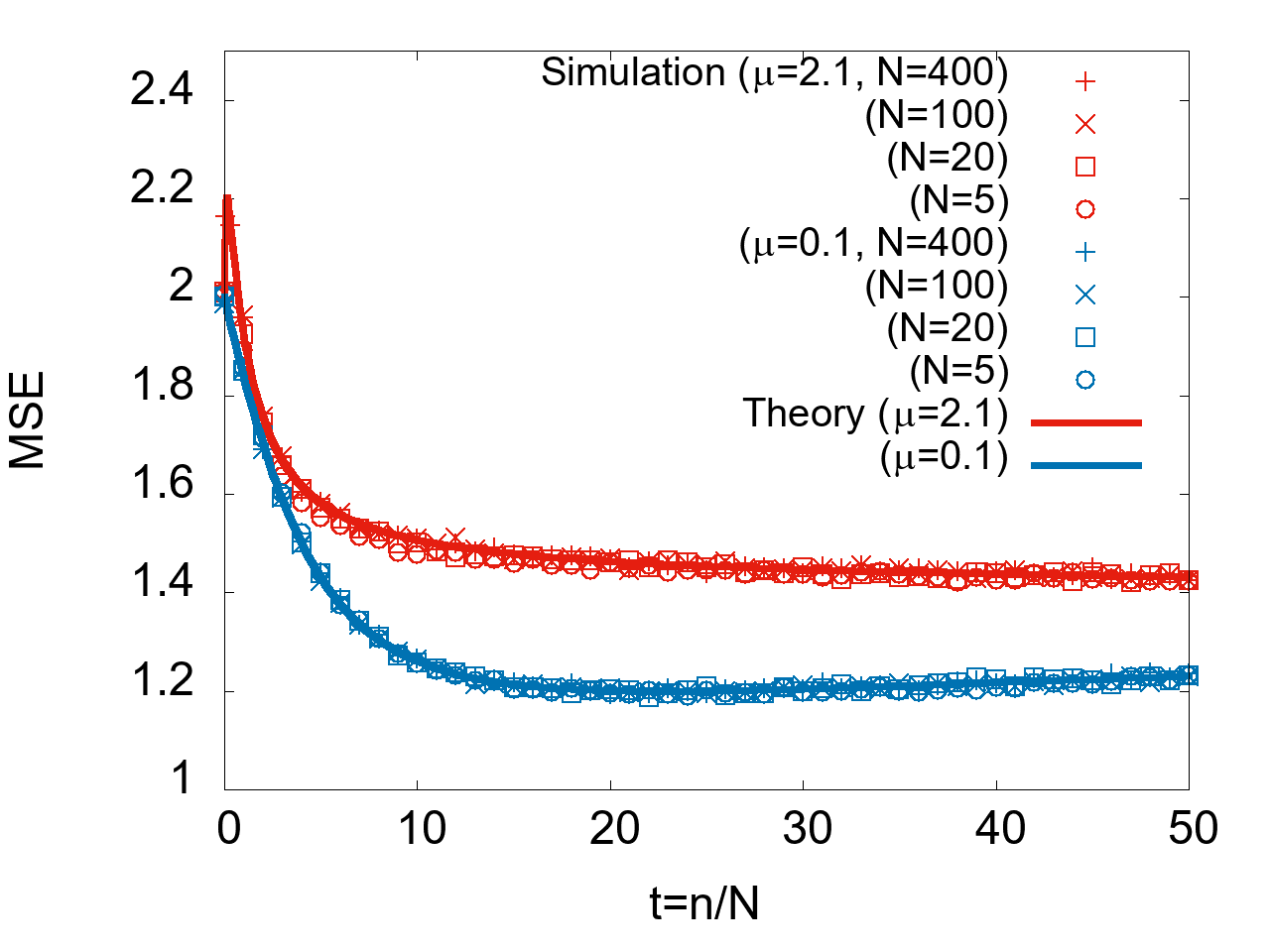}
    \subcaption{$S=1, \sigma_\xi^2=1$}\label{fig:MSES1VXI1FSE}
    \centering
	\includegraphics[width=1.00\linewidth,keepaspectratio]{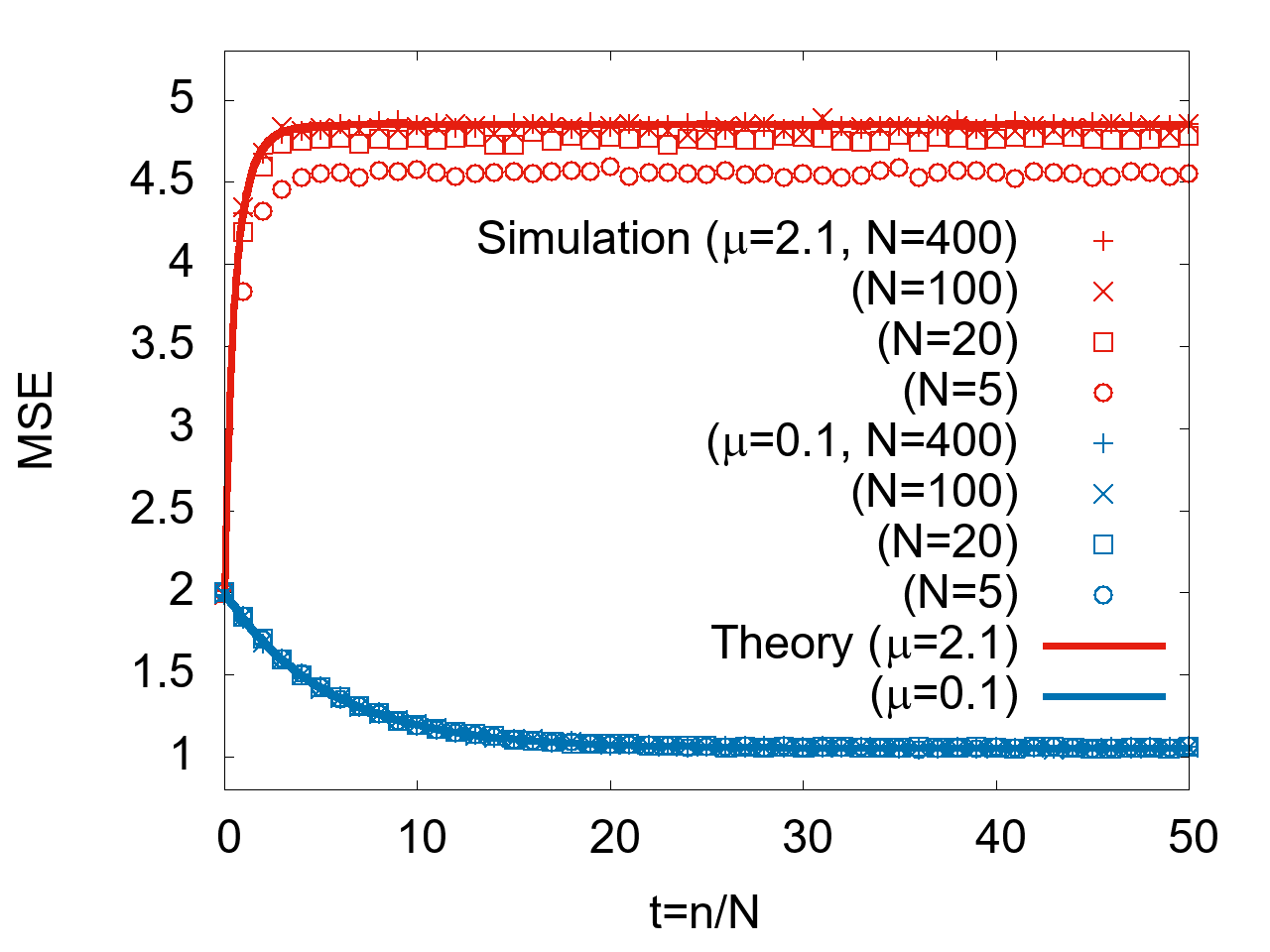}
    \subcaption{$S=3, \sigma_\xi^2=1$}\label{fig:MSES3VXI1FSE}
  \end{minipage}
%
%
  \caption{MSE learning curves obtained by simulations with small $N$ and theory.}
  \label{fig:MSEVXIFSE}
\end{figure}

Since statistical-mechanical analysis assumes $N\rightarrow \infty$, 
it is important to investigate how well the theory explains the 
simulation results with small $N$.
Figure \ref{fig:MSEVXIFSE} shows
the theoretical learning curves 
along with the simulation results
for $N=400, 100, 20$, and $5$.
In the computer simulations, the unknown system 
coefficient vectors are generated by uniformly sampling 
the impulse responses shown in Fig. \ref{fig:realunknown}
and normalizing them to satisfy (\ref{eqn:sigmag2}).
Figure \ref{fig:MSEVXIFSE} shows that, especially 
when $S$ and $\mu$ are large, there is significant disagreement
between the simulation results with small $N$ and the theoretical results.
The reason for this disagreement
is considered to be that the theory is derived 
using the long-filter assumption,
that is, $N\rightarrow \infty$,
whereas the simulations are executed using finite values of $N$.
The dependence of the disagreement on $N$
is an example of the phenomenon known as 
the {\it finite-size effect}
in statistical mechanics.
Note, however, that when $S$ or $\mu$ is small, 
the theory is in good agreement with the simulation, even if $N = 20$ or 5.

\begin{figure}[htbp]
    \centering
  \begin{minipage}[b]{0.78\linewidth}
    \centering
	\includegraphics[width=1.00\linewidth,keepaspectratio]{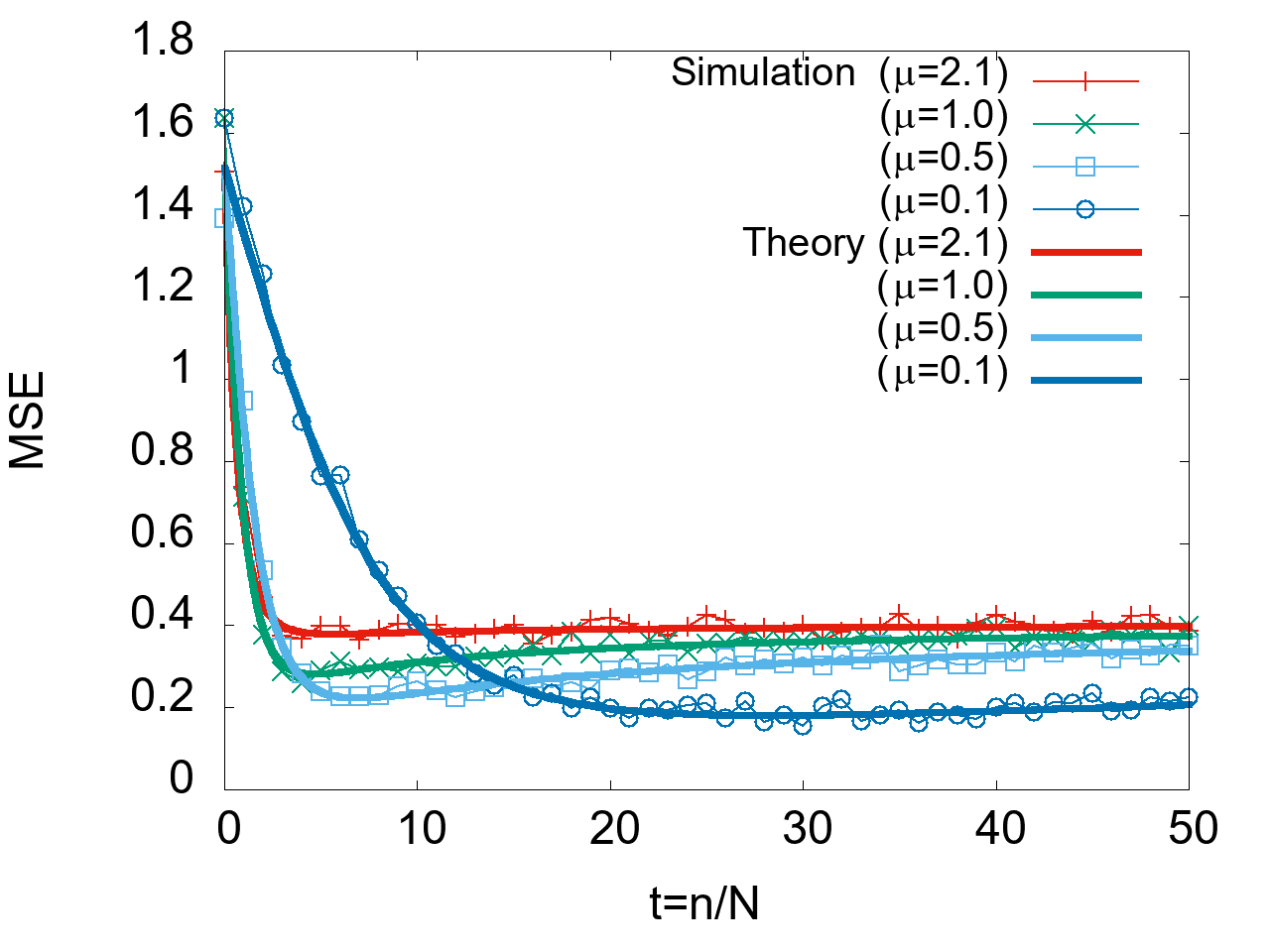}
    \subcaption{$S=1, \sigma_\xi^2=0$}\label{fig:non0w_0_S1VXI0}
    \centering
    \includegraphics[width=1.00\linewidth,keepaspectratio]{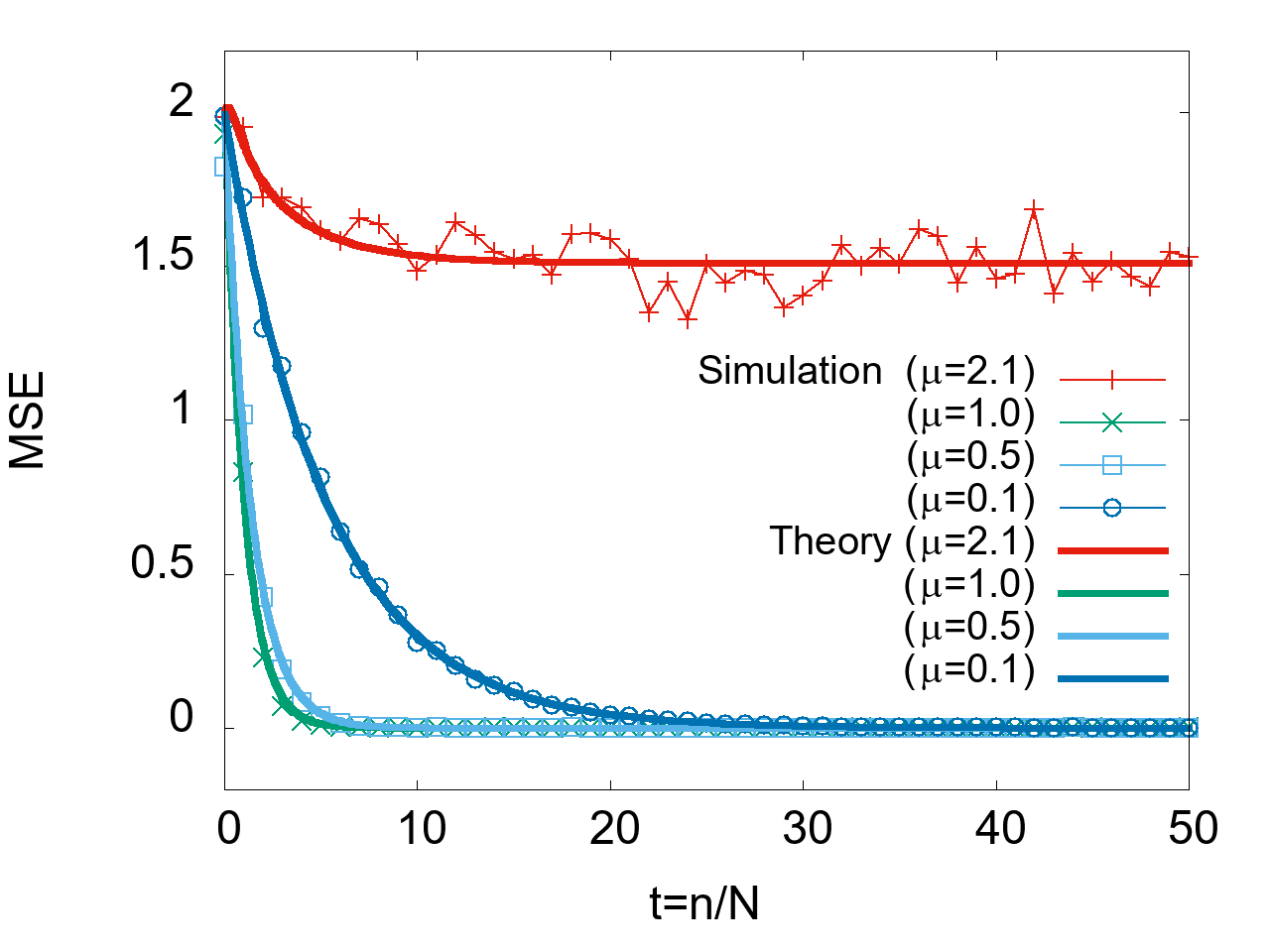}
    \subcaption{$S=3, \sigma_\xi^2=0$}\label{fig:non0w_0_S3VXI0}
  \end{minipage}
  \begin{minipage}[b]{0.78\linewidth}
    \centering
	\includegraphics[width=1.00\linewidth,keepaspectratio]{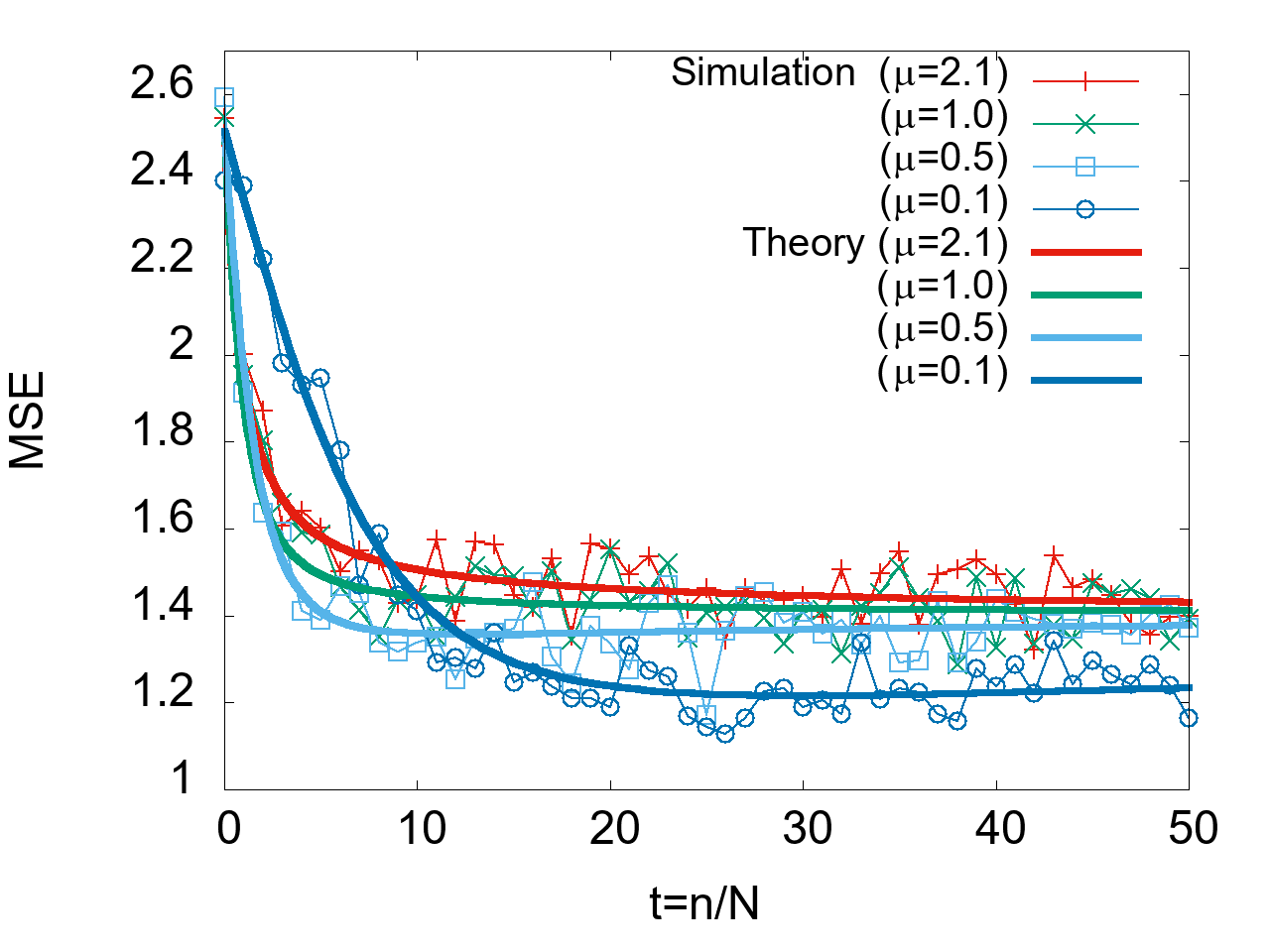}
    \subcaption{$S=1, \sigma_\xi^2=1$}\label{fig:non0w_0_S1VXI1}
    \centering
	\includegraphics[width=1.00\linewidth,keepaspectratio]{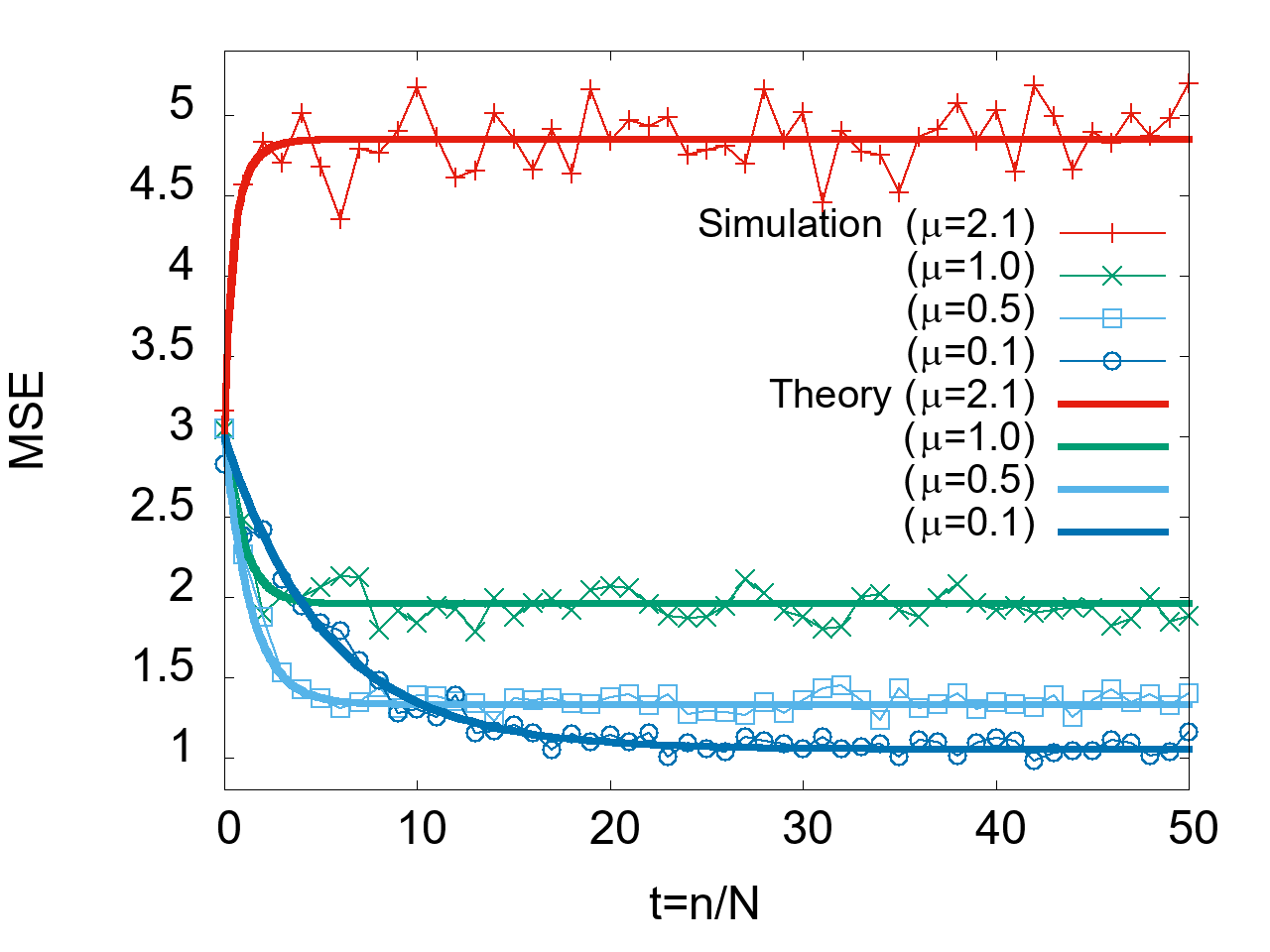}
    \subcaption{$S=3, \sigma_\xi^2=1$}\label{fig:non0w_0_S3VXI1}
  \end{minipage}
%
%
  \caption{MSE learning curves when the initial values of the filter coefficients are not zero.}
  \label{fig:non0w_0_S1,3VXI0,1}
\end{figure}

Figure \ref{fig:non0w_0_S1,3VXI0,1} shows
examples of the MSE learning curves 
when the initial values $w_i(0),\ i=1,\ldots,N$ of 
the filter coefficients are not zero.
In the computer simulation, the initial values of the filter coefficients 
are independently drawn from a Gaussian distribution 
with a mean 0 and a variance 1.
In response to this, the initial conditions $r(0)=0$ and $Q(0)=1$
are used in the theoretical calculation.
Figures \ref{fig:non0w_0_S1VXI0}--\ref{fig:non0w_0_S3VXI1} show
that the theory derived in this paper
predicts the simulation results well
even if the initial values of the filter coefficients are not zero.
Comparing these figures with 
Figs. \ref{fig:MSES1VXI0RL1}--\ref{fig:MSES3VXI1RL1},
it can be seen that the initial values of the MSE are increased 
by the non-zero initial values of the filter coefficients.
On the other hand, their steady-state values remain the same.
Hereafter, the initial values of the filter coefficients are assumed 
to be zero.

\begin{figure}[htbp]
    \centering
  \begin{minipage}[b]{0.78\linewidth}
    \centering
	\includegraphics[width=1.00\linewidth,keepaspectratio]{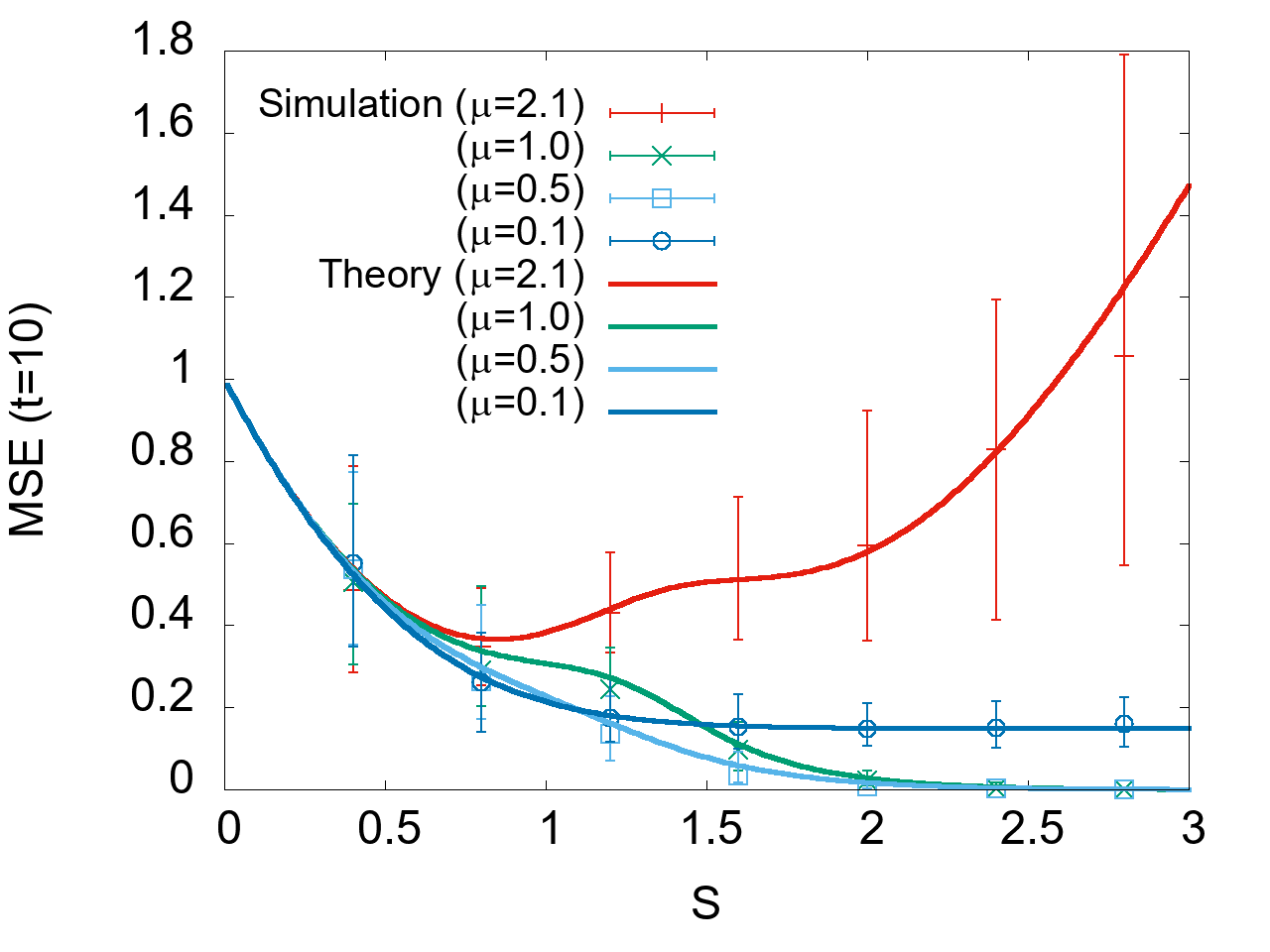}
    \subcaption{$t=10$}\label{fig:MSEVXI0TEND10RL1}
    \centering
    \includegraphics[width=1.00\linewidth,keepaspectratio]{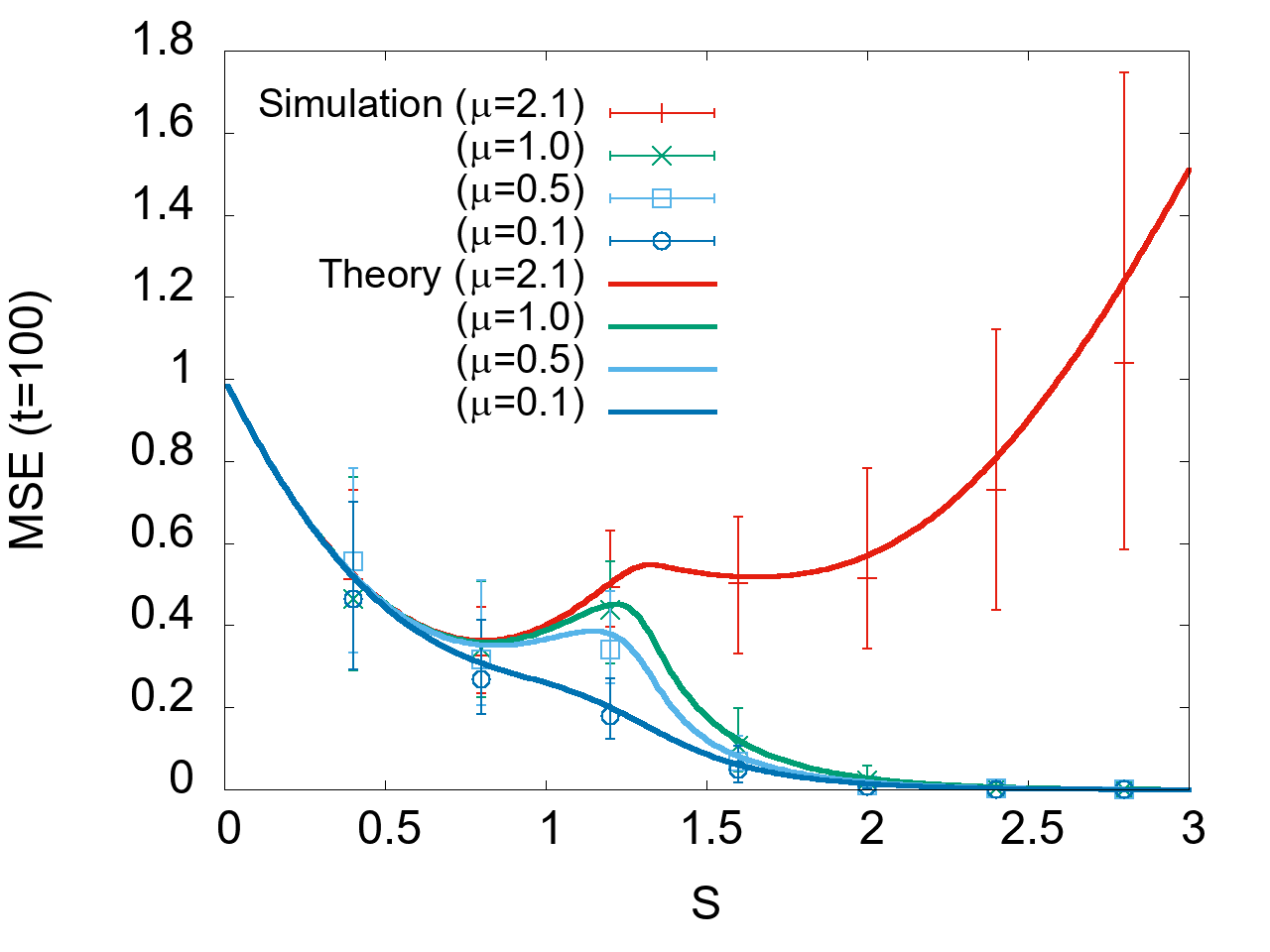}
    \subcaption{$t=100$}\label{fig:MSEVXI0TEND100RL1}
    \centering
    \includegraphics[width=1.000\linewidth,keepaspectratio]{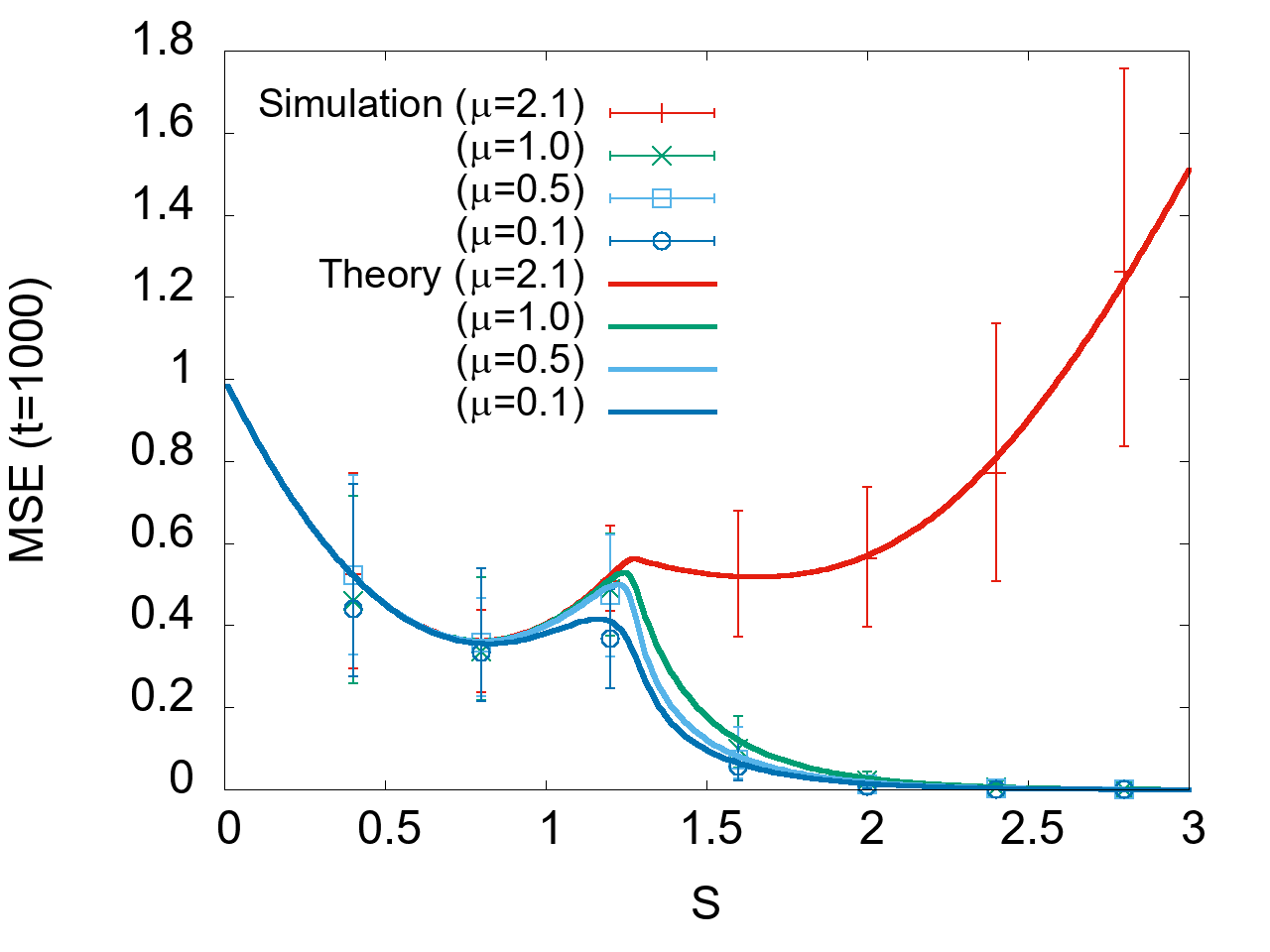}
    \subcaption{$t=1000$}\label{fig:MSEVXI0TEND1000RL1}
  \end{minipage}
%
  \caption{MSE at $t=10, 100$, and $1000$ ($\sigma_\xi^2=0$).}
  \label{fig:MSEVXI0TEND101001000RL1}
\end{figure}

\begin{figure}[htbp]
    \centering
  \begin{minipage}[b]{0.78\linewidth}
    \centering
	\includegraphics[width=1.00\linewidth,keepaspectratio]{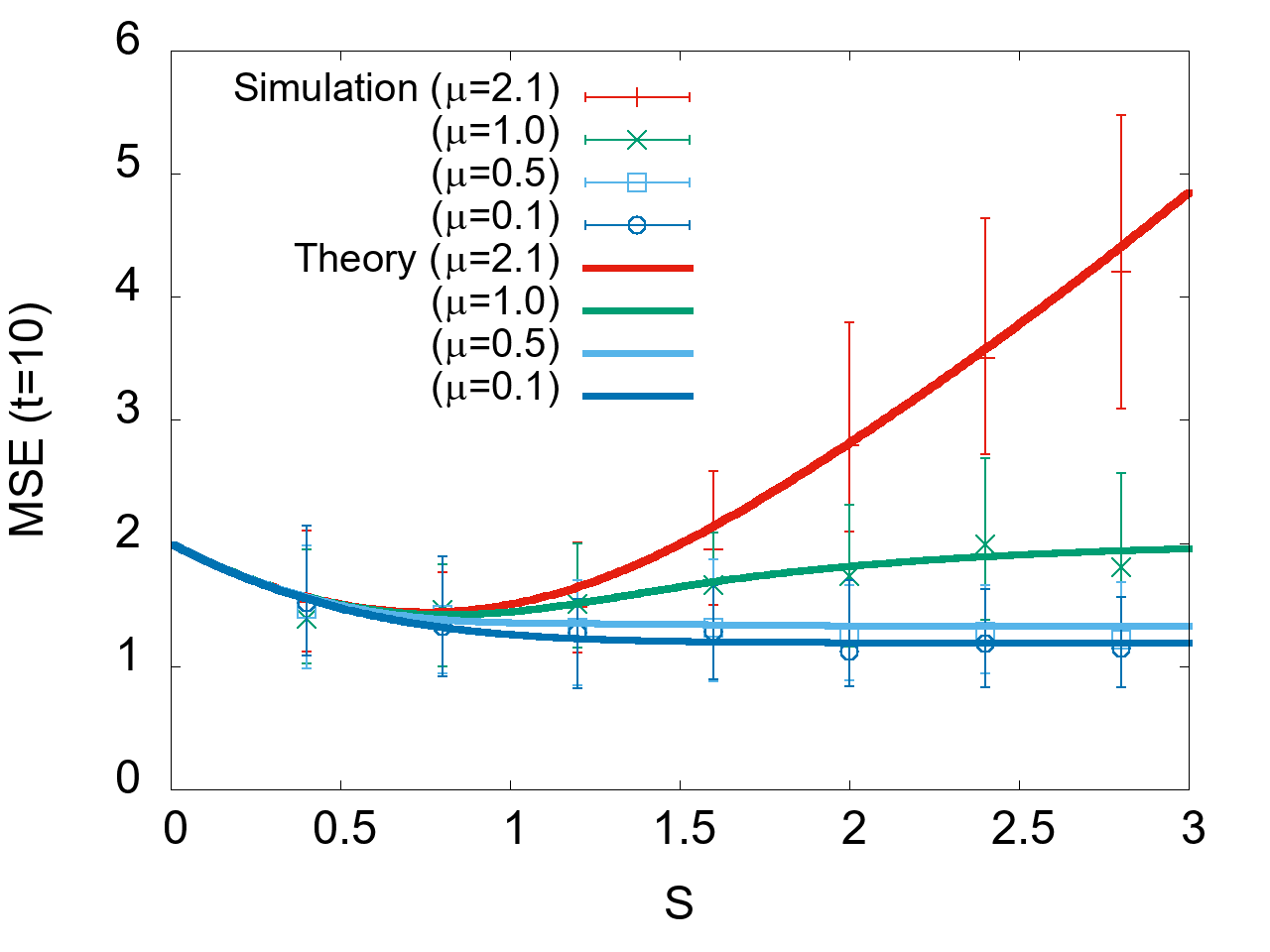}
    \subcaption{$t=10$}\label{fig:MSEVXI1TEND10RL1}
    \centering
    \includegraphics[width=1.00\linewidth,keepaspectratio]{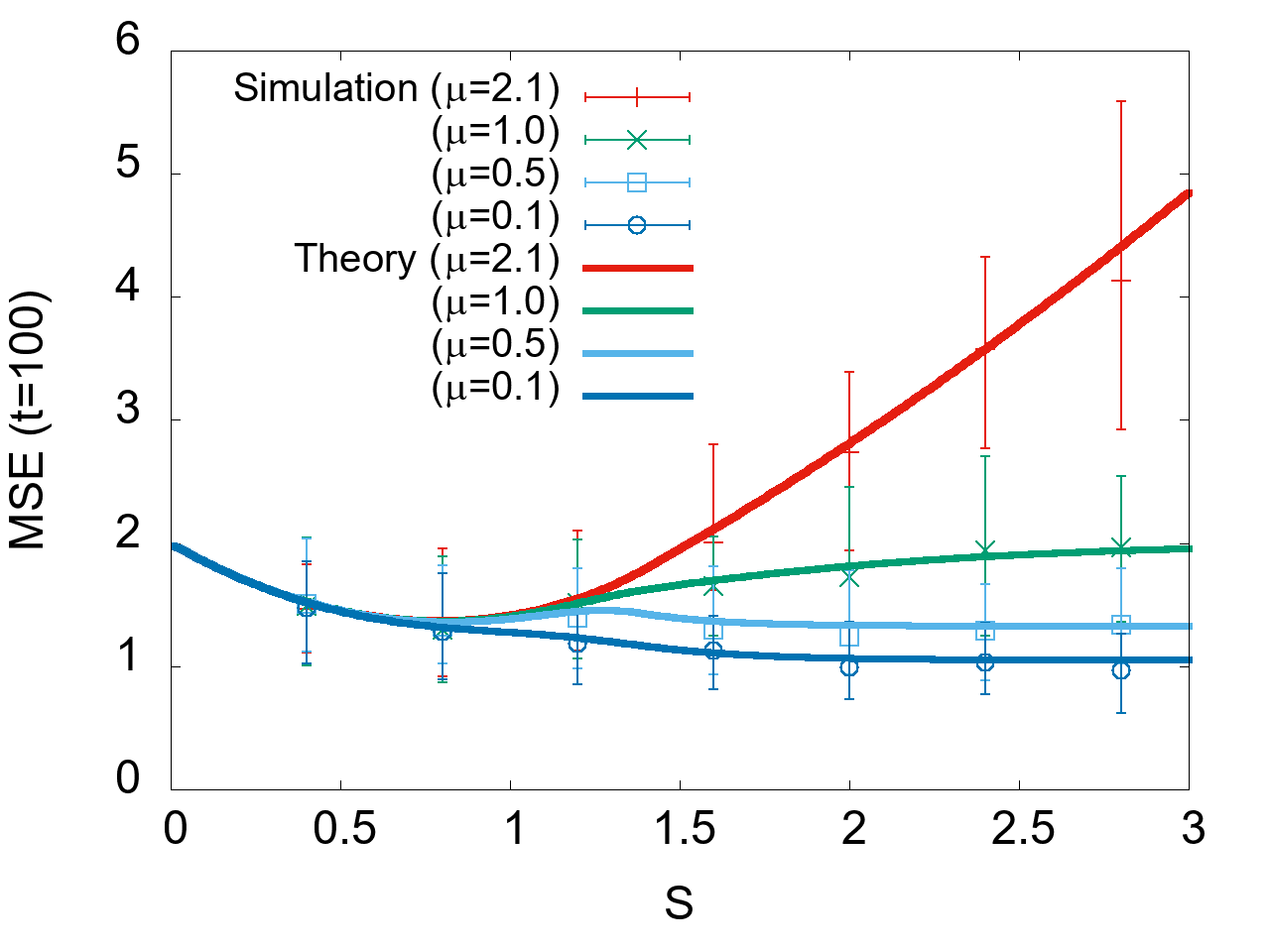}
    \subcaption{$t=100$}\label{fig:MSEVXI1TEND100RL1}
    \centering
    \includegraphics[width=1.000\linewidth,keepaspectratio]{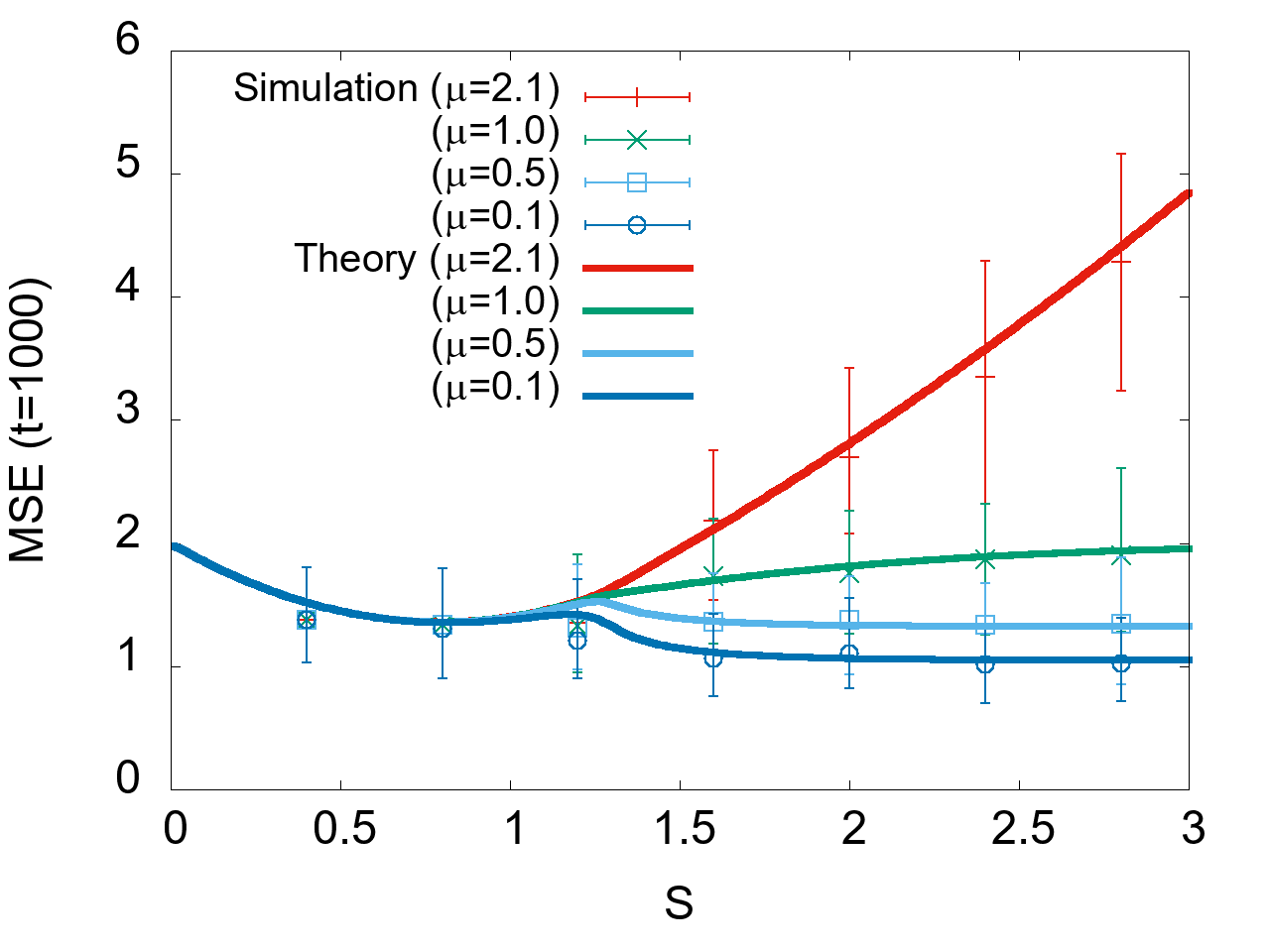}
    \subcaption{$t=1000$}\label{fig:MSEVXI1TEND1000RL1}
  \end{minipage}
%
%
  \caption{MSE at $t=10, 100$, and $1000$ ($\sigma_\xi^2=1$).}
  \label{fig:MSEVXI1TEND101001000RL1}
\end{figure}

From Figs. \ref{fig:MSES1VXI0RL1}--\ref{fig:MSES3VXI1RL1},
it seems that the MSE almost converges
at $t=50$ regardless of the step size $\mu$                 
for both $S=1$ and 3.
However, 
Figs. \ref{fig:MSDS1VXI0RL1} and \ref{fig:MSDS1VXI1RL1} show 
that 
the normalized MSD continues to increase for $S=1$.
Next, we show 
the MSE at $t=10, 100$, and $1000$
in Figs. \ref{fig:MSEVXI0TEND101001000RL1} and \ref{fig:MSEVXI1TEND101001000RL1} 
to investigate the relationship between the saturation value $S$
and the MSE.
For the computer simulations, 
the medians and standard deviations
in 100 trials are plotted using error bars.
Figures \ref{fig:MSEVXI0TEND101001000RL1} and \ref{fig:MSEVXI1TEND101001000RL1} show that 
the MSE increases 
when $S$ is in the range of 1.1--1.3, and this tendency
becomes increasingly pronounced with time.

\subsection{Critical value $S_C$ and steady-state analysis when $S>S_C$} \label{sec:steady_state}
\begin{figure}[htbp]
	\centering
  	\begin{minipage}[b]{0.78\linewidth}
    		\centering
		\includegraphics[width=1.00\linewidth,keepaspectratio]{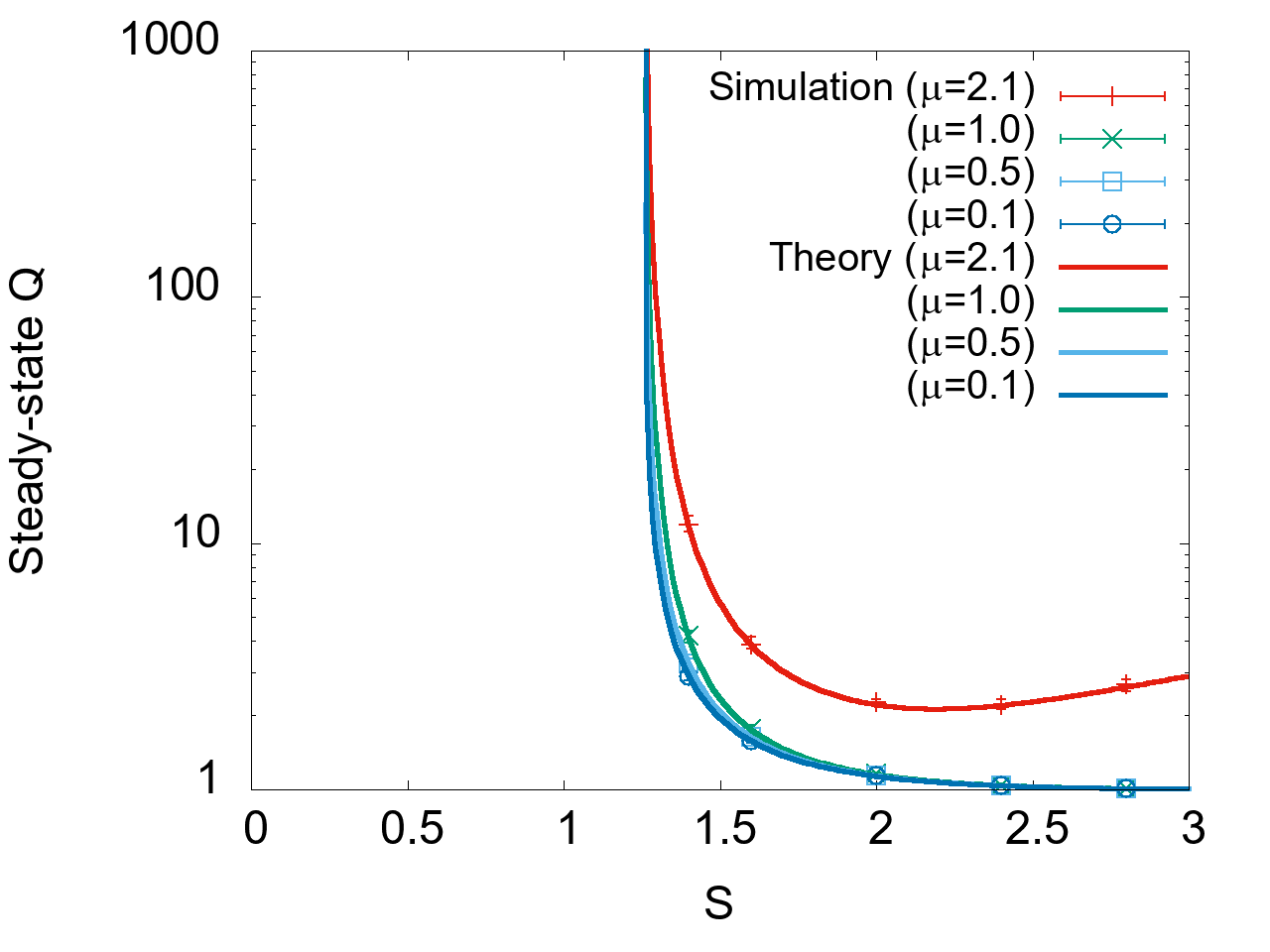}
		\subcaption{$\sigma_\xi^2=0$}\label{fig:QVXI0}
	\end{minipage}
  	\begin{minipage}[b]{0.78\linewidth}
    		\centering
		\includegraphics[width=1.00\linewidth,keepaspectratio]{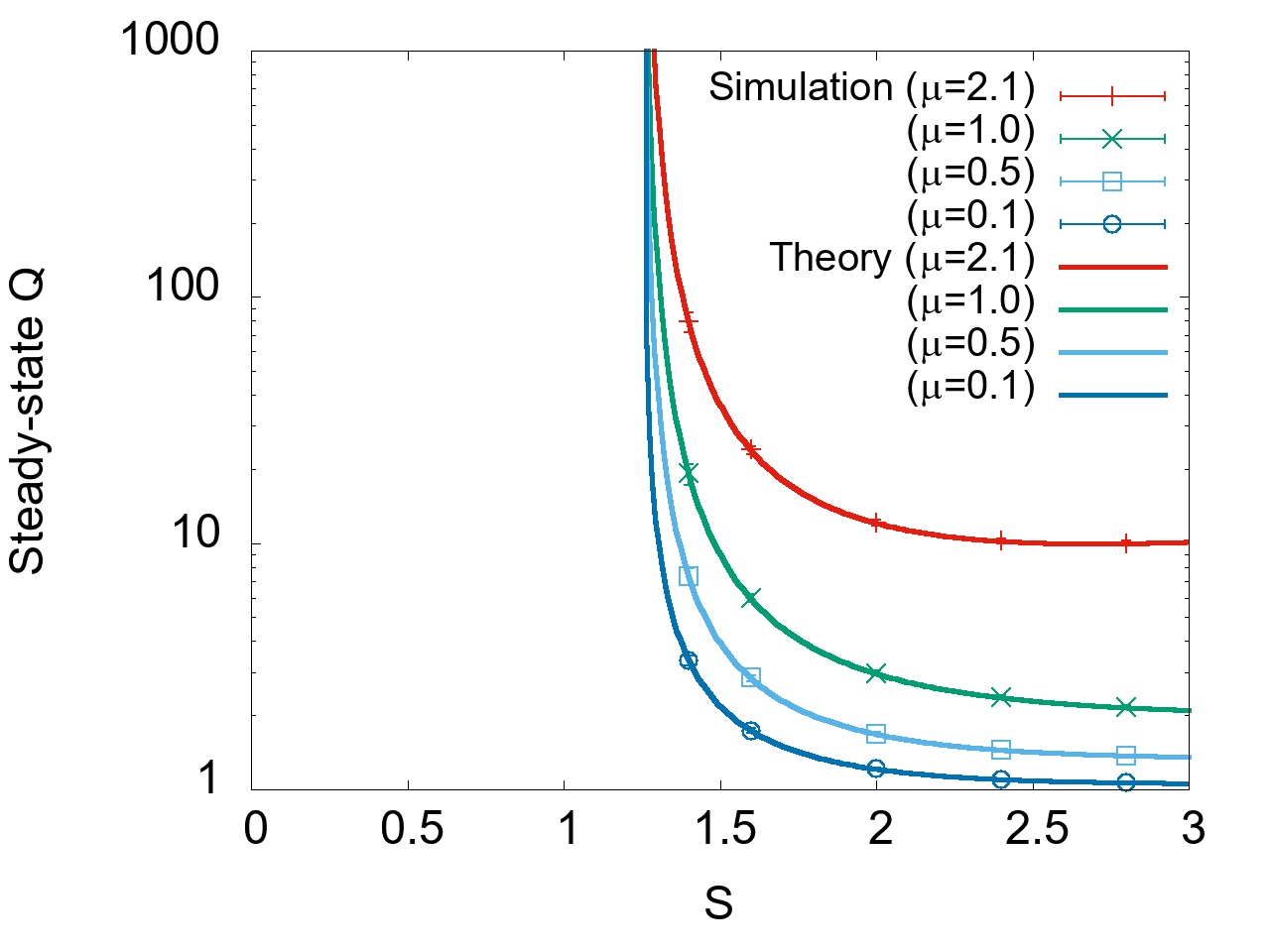}
		\subcaption{$\sigma_\xi^2=1$}\label{fig:QVXI1}
	\end{minipage}
%
%
	\caption{Steady-state $Q$.}\label{fig:QVXI0,1}
\end{figure}

\begin{figure}[htbp]
	\centering
  	\begin{minipage}[b]{0.78\linewidth}
    		\centering
		\includegraphics[width=1.00\linewidth,keepaspectratio]{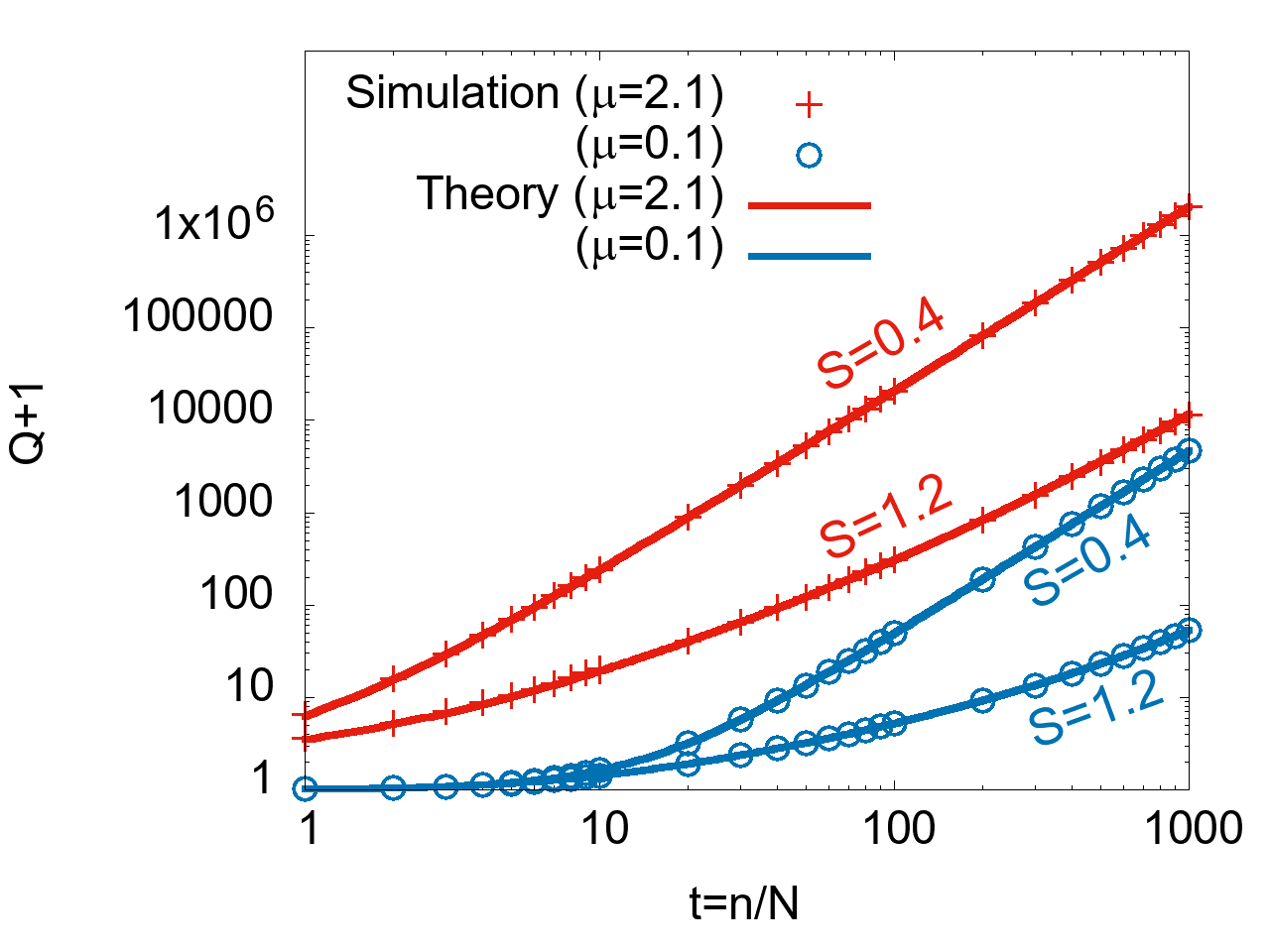}
		\subcaption{$\sigma_\xi^2=0$}\label{fig:QdivVXI0}
	\end{minipage}
  	\begin{minipage}[b]{0.78\linewidth}
    		\centering
		\includegraphics[width=1.00\linewidth,keepaspectratio]{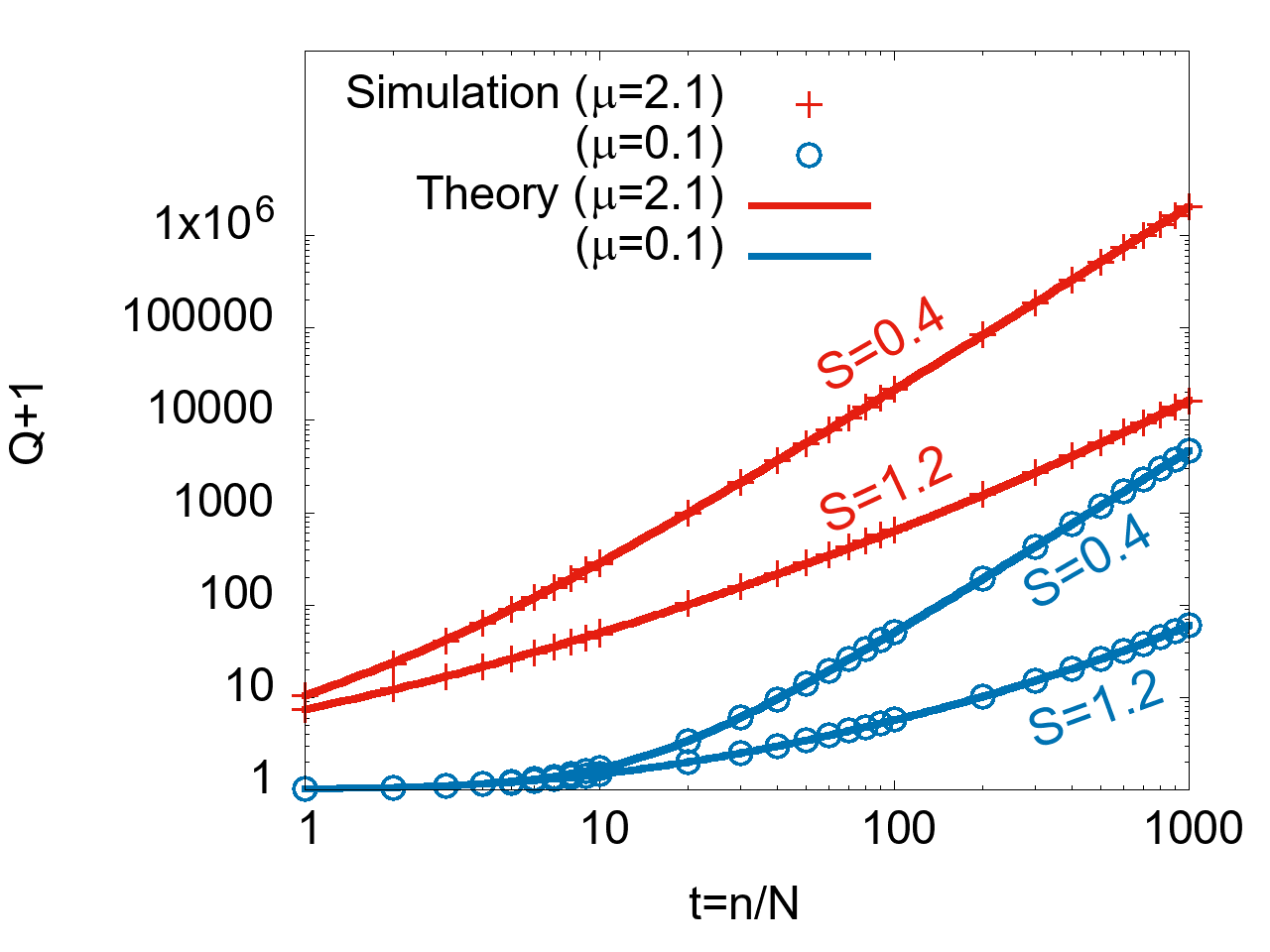}
		\subcaption{$\sigma_\xi^2=1$}\label{fig:QdivVXI1}
	\end{minipage}
%
%
	\caption{Dynamical behaviors of $Q$ when $S<S_C$.}\label{fig:QdivVXI0,1}
\end{figure}

To clarify the phenomenon
of increasing MSE described in the previous subsection,
we investigate the steady-state values of 
the macroscopic variables $r, Q$, and MSE.
If steady-state values of $r$ and $Q$ exist,
they can be obtained by 
numerically solving the simultaneous equations
that are obtained by 
substituting zeros into the left-hand sides
of the simultaneous differential equations (\ref{eqn:drdt2}) and (\ref{eqn:dQdt2}).
Figure \ref{fig:QVXI0,1} 
shows the numerically obtained results for $Q$
along with the corresponding simulation results at $t=10^4$. 
This value of $t$ in the simulation is sufficient for $Q$ to reach a steady-state value.
When the saturation value $S$ is larger than $S_C=1.25331\cdots$,
a numerical solution is found.
However, when $S$ is smaller than $S_C$,
no numerical solution is found.
Of course, no numerical solution found for the simultaneous equations  
is only a necessary condition for $r$ and $Q$ to diverge, 
not a sufficient condition. 
In other words, no numerical solution found 
does not necessarily indicate that $r$ or $Q$ diverges.
Of course, it is most desirable 
to prove mathematically that $r$ and/or $Q$ diverge, 
but this is difficult because of the complexity of 
the simultaneous differential equations (\ref{eqn:drdt2}) and (\ref{eqn:dQdt2}).
Therefore, we investigate the dynamical behavior of $Q$ 
for $S<S_C$ by means of theoretical numerical calculations 
and computer simulations.
Figures \ref{fig:QdivVXI0} and \ref{fig:QdivVXI1} show the results.
In these figures, $Q$ increases monotonically in all cases. 
These results strongly indicate that 
$Q$ diverges regardless of 
the step size and the presence of background noise when $S<S_C$. 
Since $Q$ is proportional to the square of the $\ell_2$-norm of $\bm{w}$, as 
seen from (\ref{eqn:Qdef}),
the divergence of $Q$ means the divergence of 
the coefficient vector $\bm{w}$ of the adaptive filter W.
When $S>S_C$, we can obtain the steady-state MSE by substituting 
the steady-state values of $r$ and $Q$ into (\ref{eqn:MSE2}).
Note that it is clarified in Sec. \ref{sec:S_C}
that the exact value of $S_C=1.25331\cdots$
is $\sqrt{\frac{\pi}{2}}$.

\subsection{Asymptotic analysis when $S<S_C$} \label{sec:asym}
On the basis of the results obtained in the previous subsection, 
we henceforth assume that 
$Q$ diverges in the limit $t\rightarrow\infty$ when $S<S_C$.
However, Figs. \ref{fig:MSEVXI0TEND101001000RL1} 
and \ref{fig:MSEVXI1TEND101001000RL1}
show that the MSE appears to converge even when $S<S_C$.
Therefore, an asymptotic analysis of  
the behavior of the system when $S<S_C$ 
is given in this subsection.
From (\ref{eqn:sigmag2}), (\ref{eqn:Qdef}), and (\ref{eqn:rdef}), we obtain
\begin{align}
r&= \sigma_g \sqrt{Q} \cos \theta. \label{eqn:Qr}
\end{align}
Here, $\theta$ is the angle between vectors $\bm{g}$ and $\bm{w}$.
From (\ref{eqn:drdt2}), (\ref{eqn:dQdt2}), and (\ref{eqn:Qr}),
we obtain
\begin{align}
\cos \theta &\xrightarrow{t\rightarrow \infty}1. \label{eqn:R=1}
\end{align}
Equation (\ref{eqn:R=1}) is derived in detail in Appendix \ref{sec:appR=1},
where the approximations
\begin{align}
\mbox{erf}\left(x \right)
&=\frac{2}{\sqrt{\pi}}\sum_{n=0}^\infty \frac{(-1)^n x^{2n+1}}{n! (2n+1)} 
\simeq \frac{2}{\sqrt{\pi}}x &(|x|\ll 1),
\label{eqn:erf_approx} \\
\exp\left(x \right)
&=\sum_{n=0}^\infty \frac{x^n}{n!} \simeq 1+x &(|x|\ll 1)
\label{eqn:exp_approx}
\end{align}
are used.

Equation (\ref{eqn:R=1}) means that the directions of $\bm{g}$ and $\bm{w}$
coincide even when $S<S_C$,
although $\bm{w}$ diverges 
as described in Sec. \ref{sec:steady_state}.
Note that this property does not depend on 
$\mu, \rho^2,\sigma_g^2$, or $\sigma_\xi^2$.

\begin{figure}[htbp]
	\centering
  	\begin{minipage}[b]{0.78\linewidth}
    		\centering
		\includegraphics[width=1.00\linewidth,keepaspectratio]{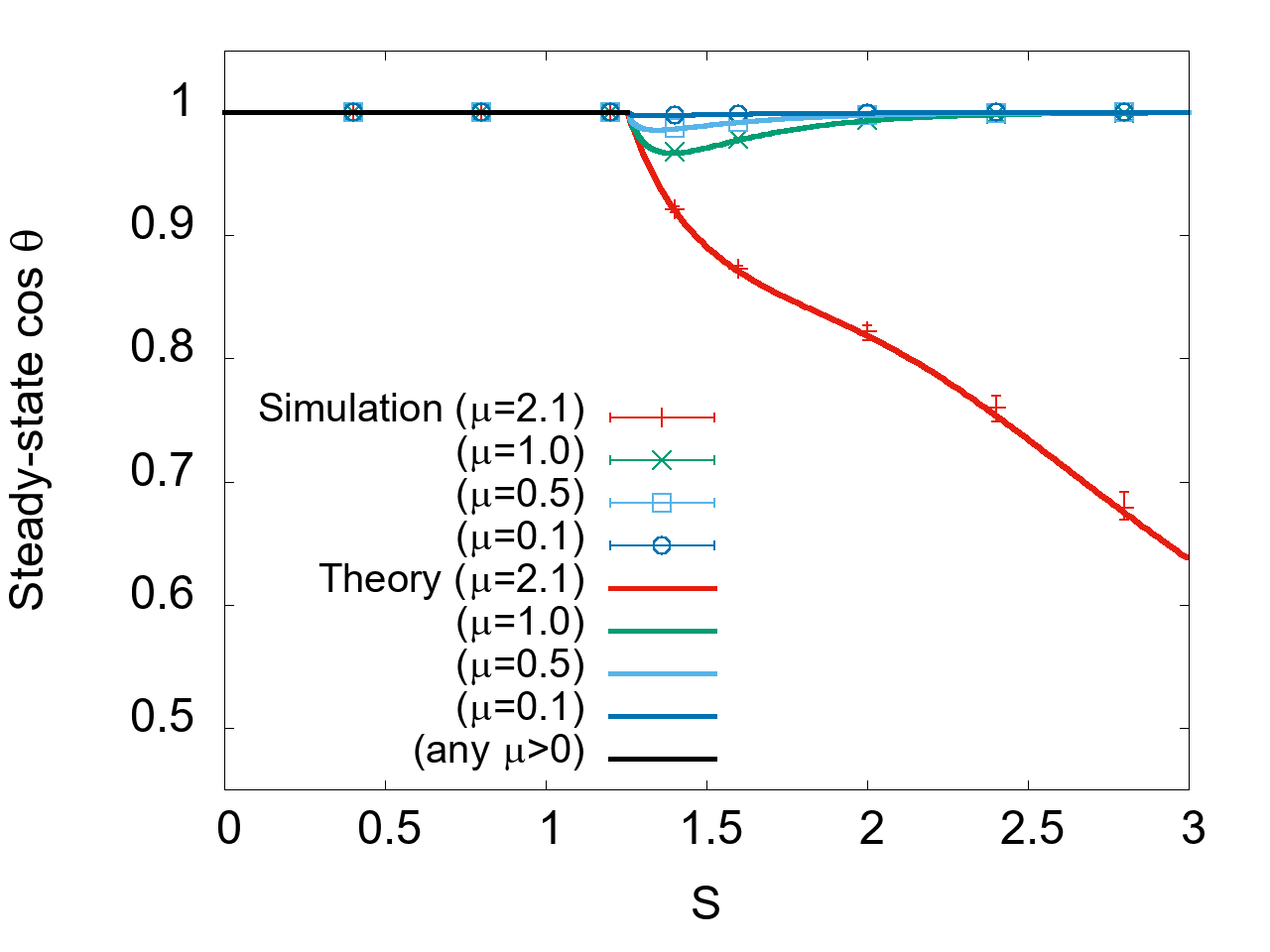}
		\subcaption{$\sigma_\xi^2=0$}\label{fig:RVXI0}
	\end{minipage}
  	\begin{minipage}[b]{0.78\linewidth}
    		\centering
		\includegraphics[width=1.00\linewidth,keepaspectratio]{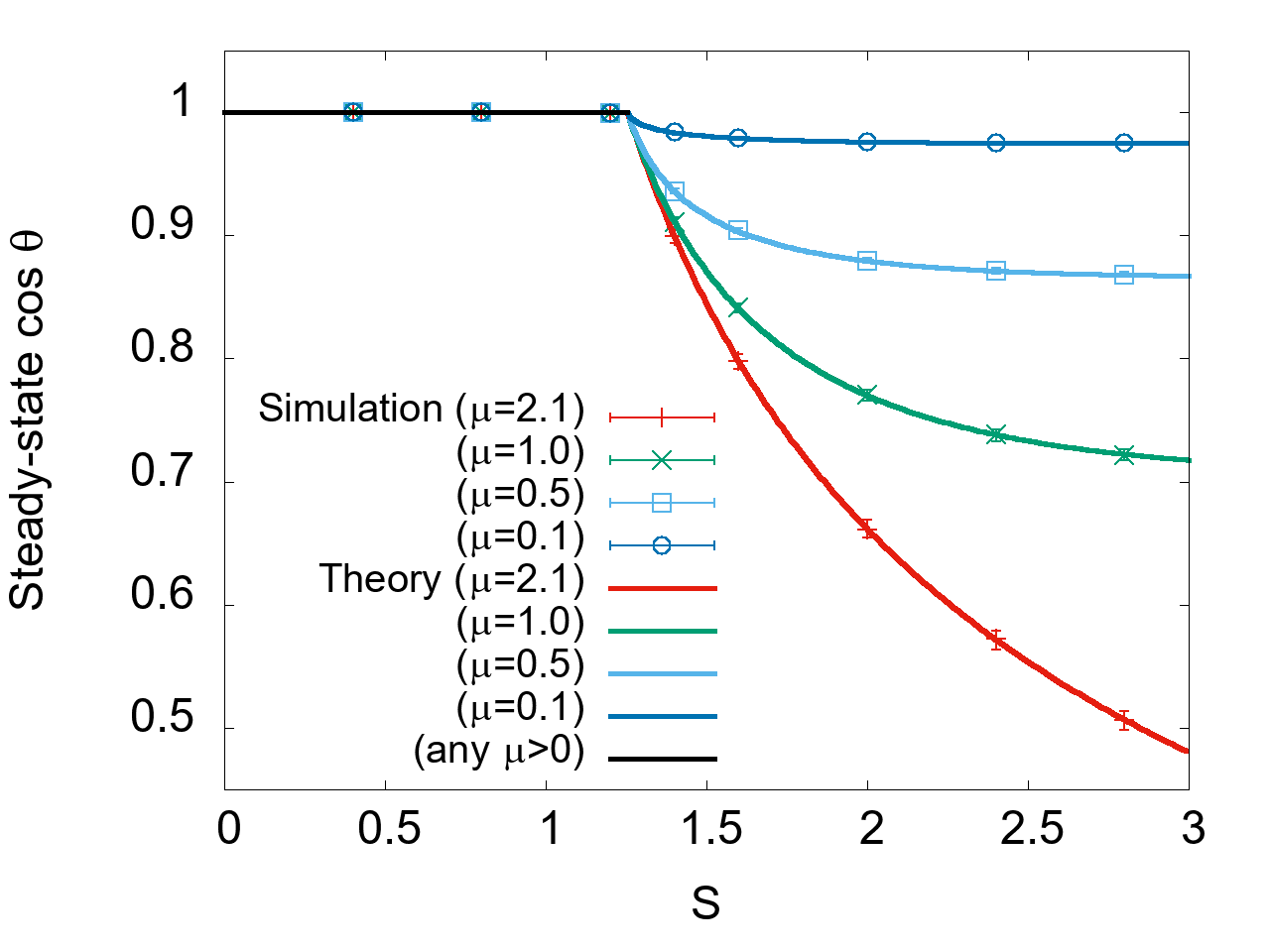}
		\subcaption{$\sigma_\xi^2=1$}\label{fig:RVXI1}
	\end{minipage}
%
%
	\caption{Steady-state $\cos \theta=r/(\sigma_g \sqrt{Q})$.}\label{fig:RVXI0,1}
\end{figure}

As described so far,
$Q$ diverges and $\cos \theta =1$ in the limit $t\rightarrow\infty$ when $S<S_C$.
Then, the MSE is
\begin{align}
\lla e^2 \rra
&=
S^2-2\sigma_g \rho \sqrt{\frac{2}{\pi}} S 
+ \sigma_g^2 \rho^2 + \sigma_\xi^2.
\label{eqn:MSE3}
\end{align}
Equation (\ref{eqn:MSE3}) is derived in detail in Appendix \ref{sec:appMSE3}.
Although $\bm{w}$ diverges when $S<S_C$,
(\ref{eqn:MSE3}) shows that the MSE converges.
The converged value does not depend on the step size $\mu$,
and it is a quadratic function of $S$.
The converged value is minimum, 
$\sigma_g^2 \rho^2 \lp 1-\frac{2}{\pi}\rp + \sigma_\xi^2$,
when $S= \sigma_g \rho \sqrt{\frac{2}{\pi}}$.
Figure \ref{fig:RVXI0,1} shows the theoretically obtained steady-state values
of $\cos \theta=r/(\sigma_g \sqrt{Q})$,
along with the corresponding simulation results at $t=10^4$. 
This value of $t$ in the simulation is sufficient for $\cos \theta$ to 
reach a steady-state value.
In this figure, for $S>S_C$, 
theoretically obtained values calculated using the steady-state values of $r$ and $Q$
given in Sec. \ref{sec:steady_state}
are plotted.
On the other hand, 
for $S<S_C$, 
the theoretically obtained value obtained using (\ref{eqn:R=1}) is plotted.



\begin{figure}[htbp]
    \centering
  \begin{minipage}[b]{0.78\linewidth}
    \centering
	\includegraphics[width=1.00\linewidth,keepaspectratio]{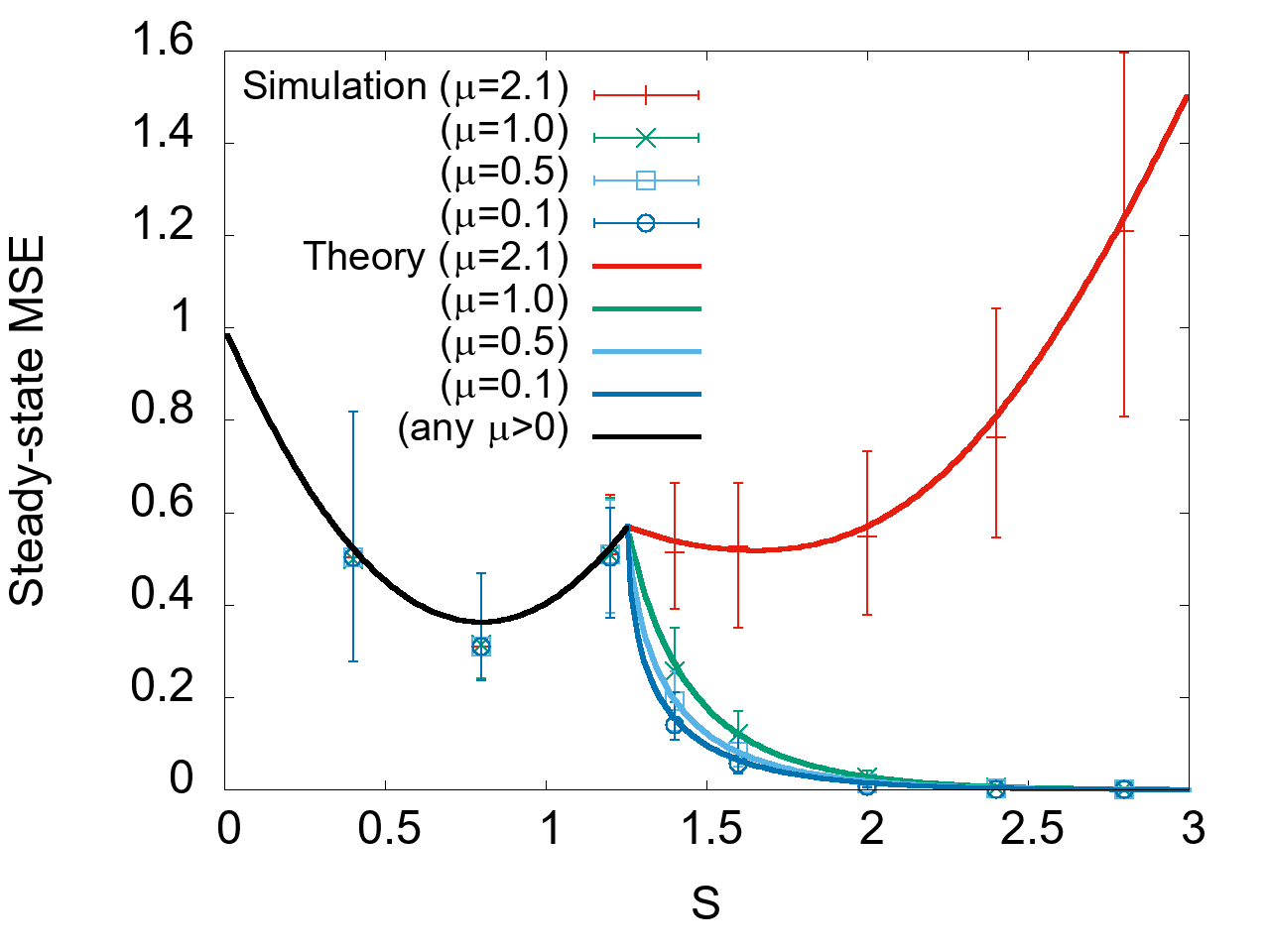}
    \subcaption{$\sigma_\xi^2=0$}\label{fig:MSEsteadystateVXI0}
  \end{minipage}
  \begin{minipage}[b]{0.78\linewidth}
    \centering
	\includegraphics[width=1.00\linewidth,keepaspectratio]{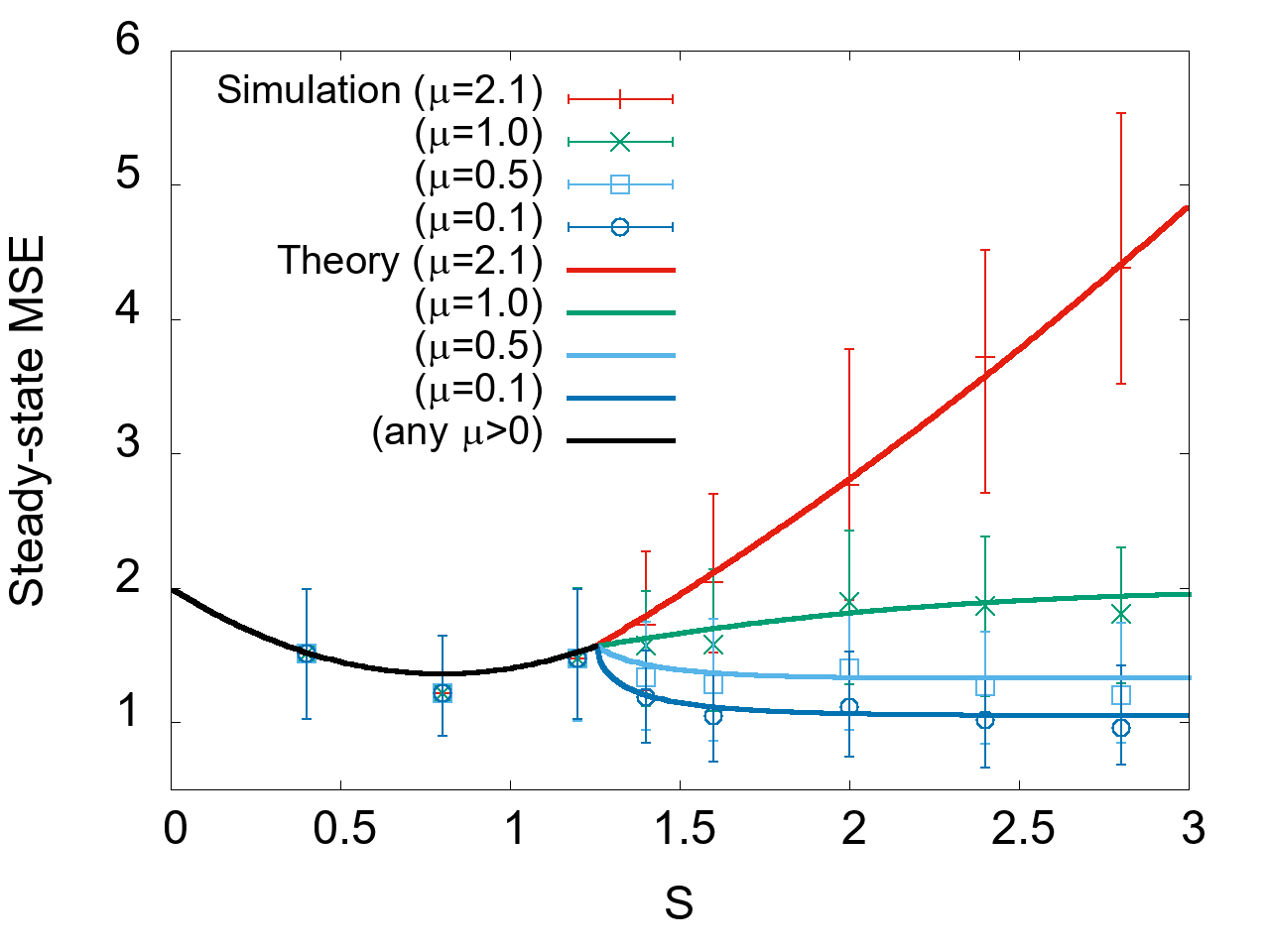}
    \subcaption{$\sigma_\xi^2=1$}\label{fig:MSEsteadystateVXI1}
  \end{minipage}
%
%
  \caption{Steady-state MSE.}\label{fig:MSEsteadystate}
\end{figure}

Figure \ref{fig:MSEsteadystate} shows 
the theoretically obtained steady-state values of the MSE,
along with the corresponding simulation results at $t=10^4$. 
This value of $t$ in the simulation is sufficient for the MSE to reach a steady-state value.
In this figure, for $S>S_C$,
the theoretically obtained results of the steady-state analysis described in 
Sec. \ref{sec:steady_state} are plotted.
On the other hand, for $S<S_C$,
the theoretical result obtained using (\ref{eqn:MSE3}) is plotted.
As 
Fig. \ref{fig:MSEsteadystate}  
and (\ref{eqn:MSE3}) show, 
for $S<S_C$, the steady-state MSE is independent of the step size $\mu$.
On the other hand, for $S>S_C$, the steady-state MSE depends on 
$\mu$, 
but the MSE never diverges because 
$|f((y(n))|\leq S$.
From the above, there is no upper bound on $\mu$ 
for the MSE to converge.

\begin{figure}[htbp]
    \centering
  \begin{minipage}[b]{0.78\linewidth}
    \centering
	\includegraphics[width=1.00\linewidth,keepaspectratio]{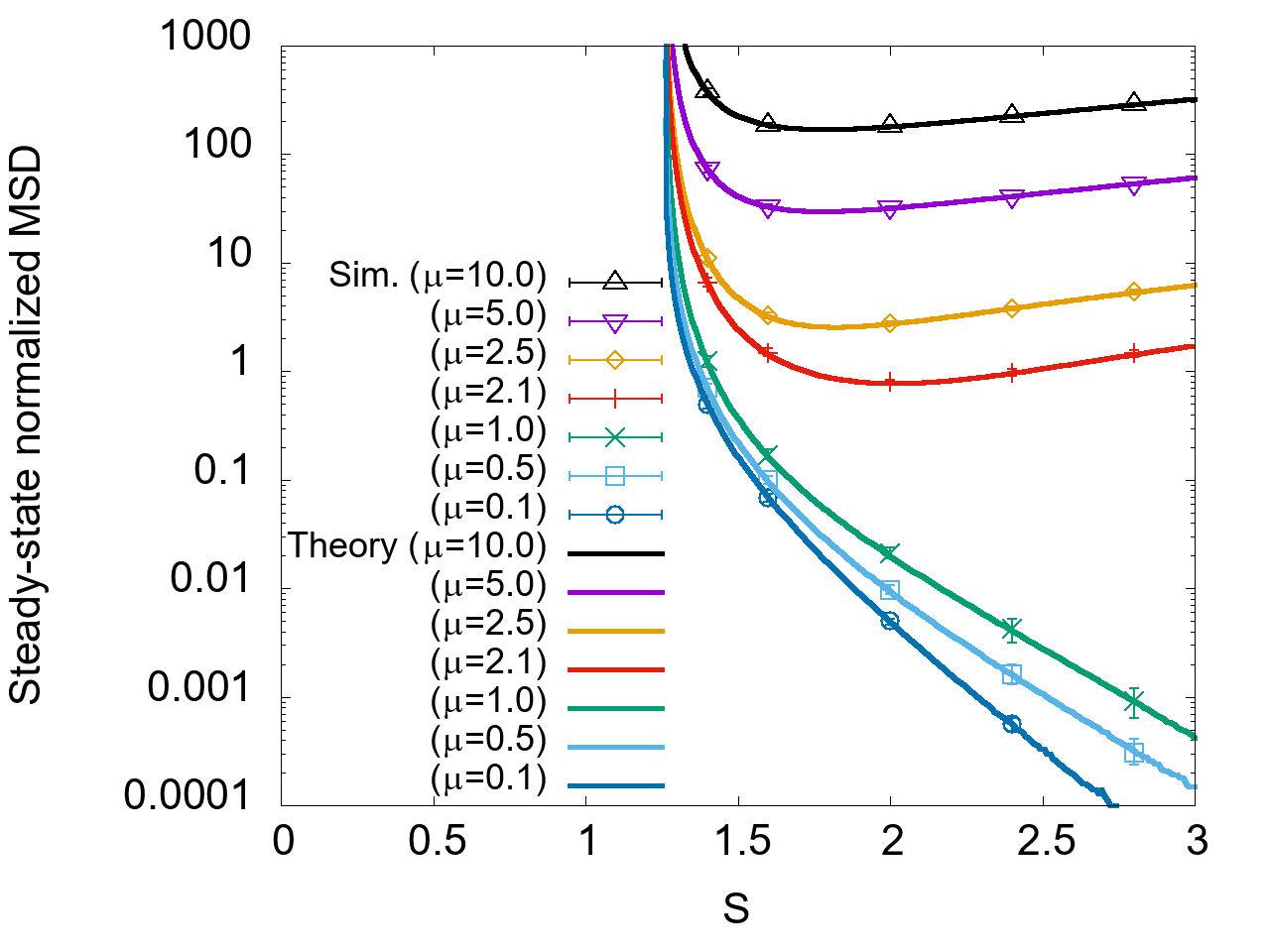}
    \subcaption{$\sigma_\xi^2=0$}\label{fig:MSDsteadystateVXI0}
  \end{minipage}
  \begin{minipage}[b]{0.78\linewidth}
    \centering
	\includegraphics[width=1.00\linewidth,keepaspectratio]{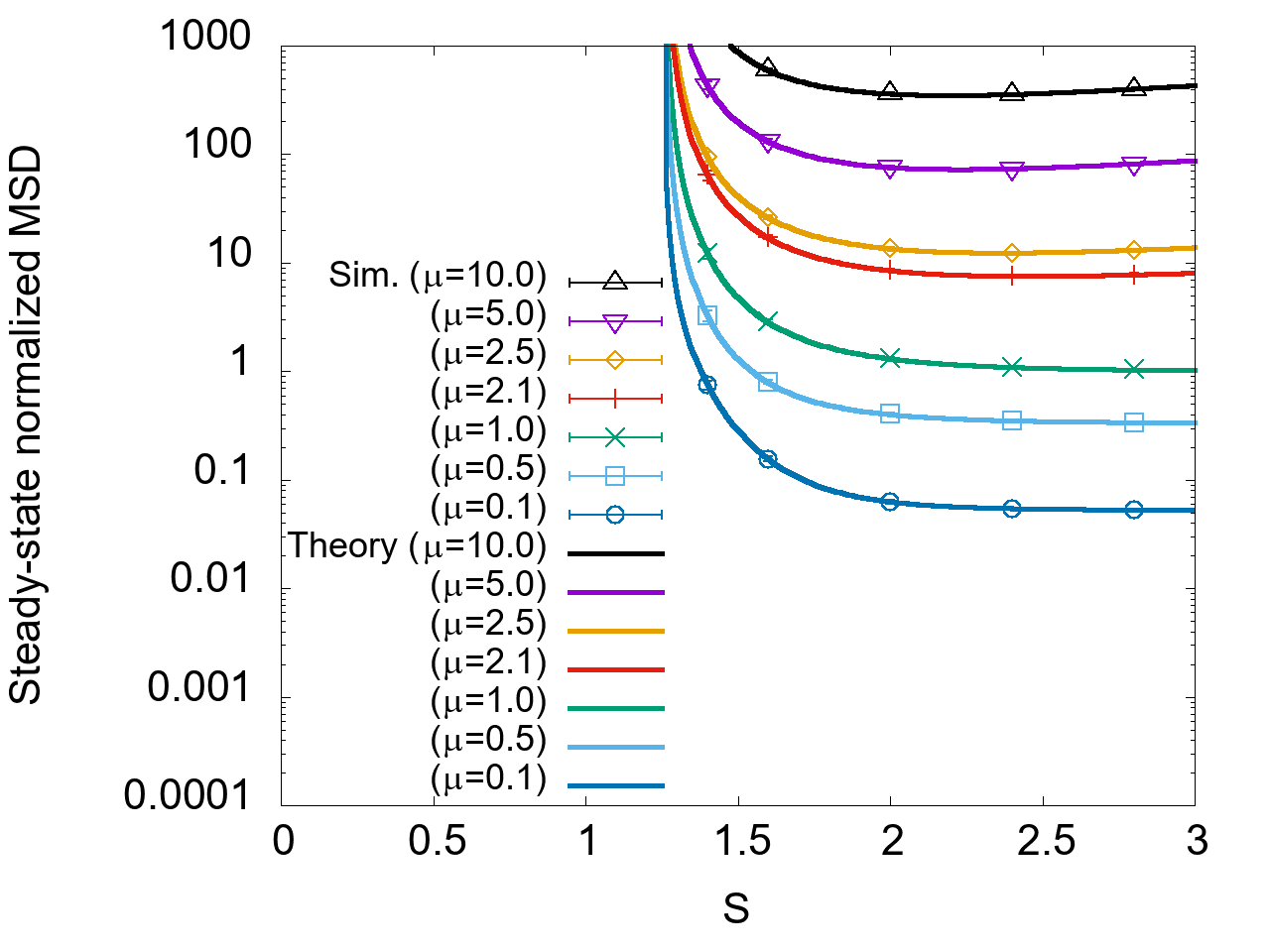}
    \subcaption{$\sigma_\xi^2=1$}\label{fig:MSDsteadystateVXI1}
  \end{minipage}
%
%
  \caption{Steady-state normalized MSD.}\label{fig:MSDsteadystate}
\end{figure}

Figure \ref{fig:MSDsteadystate} shows 
the theoretically obtained steady-state normalized MSD,
along with the corresponding simulation results at $t=10^4$. 
This value of $t$ in the simulation is sufficient for the normalized MSD 
to reach a steady-state value.
This figure shows 
that the normalized MSD diverges
because $Q$ diverges in the limit 
$S\downarrow S_C$,
where $S\downarrow S_C$ denotes that $S$ decreases in value 
approaching $S_C$.
That is, the condition in which the adaptive system is 
mean-square stable\cite{Sayed2003} is $S>S_C$.
Although not mathematically proven, 
Fig. \ref{fig:MSDsteadystate} suggests 
that for $S>S_C$, the normalized MSD converges 
regardless of the step size $\mu$.

\subsection{Exact derivation of the critical value $S_C$} \label{sec:S_C}
As described so far, the properties of the adaptive system
switch markedly
at $S=S_C$.
That is, when $S>S_C$,
the MSE and normalized MSD converge.
In other words, the adaptive system is mean-square stable.
The converged values depend on the step size $\mu$.
On the other hand,
when $S<S_C$,
the normalized MSD diverges and
the MSE converges to a value that does not depend on 
$\mu$.
In addition, the angle between the coefficient vector $\bm{g}$ of 
the unknown system G and  the coefficient vector $\bm{w}$ of
the adaptive filter W converges to zero.

In this subsection,
we analytically obtain the critical value $S_C$.
As described in Sec. \ref{sec:steady_state},
\begin{align}
\lim_{S \downarrow S_C}
\lim_{t\rightarrow \infty}
\frac{\mathrm{d}r}{\mathrm{d}t}
=
\lim_{S \downarrow S_C}
\lim_{t\rightarrow \infty}
\frac{\mathrm{d}Q}{\mathrm{d}t} 
&=0 \label{eqn:drdtdQdt=0}.
\end{align}
As described in Secs. \ref{sec:steady_state} and \ref{sec:asym},
\begin{align}
\lim_{S \downarrow S_C}
\lim_{t\rightarrow \infty}
Q&= \infty, \label{eqn:Qinf}\\
\lim_{S \downarrow S_C}
\lim_{t\rightarrow \infty}
\cos \theta&=1 \label{eqn:cos1}.
\end{align}
Substituting 
(\ref{eqn:erf_approx}), (\ref{eqn:exp_approx}), 
and (\ref{eqn:drdtdQdt=0})--(\ref{eqn:cos1})
into (\ref{eqn:drdt2}) and (\ref{eqn:dQdt2})
and solving for $S_C$,
we obtain 
\begin{align}
S_C &= \sigma_g \rho \sqrt{\frac{\pi}{2}}. \label{eqn:S_C}
\end{align}
Equation (\ref{eqn:S_C}) shows that the critical value $S_C$
is proportional to 
$\sigma_g$
and
$\rho$
defined by (\ref{eqn:sigmag2}) and (\ref{eqn:rho}), respectively.
In addition, we see that the 
critical value of $S_C=1.25331\cdots$ numerically obtained 
in Sec. \ref{sec:steady_state}
is, in fact, $\sqrt{\frac{\pi}{2}}$.

\section{Conclusions} \label{sec:conclusions}
We analyzed the behaviors of an adaptive system
in which the output of the adaptive filter 
has the clipping saturation-type nonlinearity
by the statistical-mechanical method. 
As a result, it was clarified 
that the saturation value $S$ has a critical value $S_C$
at which the system's mean-square stability 
and instability switch.
Finally, $S_C$ was exactly derived by asymptotic analysis.
Although the theory by Costa {\it et al.}\cite{Costa2001} and 
the theory proposed in this study
take completely different approaches
in terms of assumptions and methods, 
they both reveal the existence of critical values 
for nonlinearity, which is extremely interesting.
That is, $\eta^2=1$ is the condition for establishing the ``power threshold'' 
in \cite{Costa2001}, and $S=S_C=\sigma_g \rho \sqrt{\frac{\pi}{2}}$ 
is the critical condition in our theory.
However, our theory is valid for arbitrary step sizes, 
whereas Costa {\it et al.}\cite{Costa2001} assume small step sizes. 
Our theory also well explains the simulation results in the case of 
strong nonlinearity $S<\sigma_g \rho \sqrt{\frac{\pi}{2}}$, 
which corresponds to $\eta^2>1$ in \cite{Costa2001}.
The findings in this study are non-trivial, albeit 
with respect to classical and simple models, 
and are significant both theoretically and practically.

Interesting theoretical problems that should be studied further remain.
Analyses of other types of nonlinearities should be performed.
For example, a dead-zone-type nonlinearity expressed by a piecewise linearity
can be analyzed by the method described in this paper.
In addition,
the nonlinearities of other components 
should be analyzed in future studies.

\section*{Acknowledgements}
The author wishes to thank Professor Yoshinobu Kajikawa
for providing the actual data of the experimentally measured impulse response.

\onecolumn

\appendices
\section{Derivation of means and variance--covariance matrix of $d$ and $y$}
\label{sec:app_mean_cov}
From (\ref{eqn:sigmag2}), 
(\ref{eqn:u_mean_variance}), 
(\ref{eqn:d}), (\ref{eqn:y}), (\ref{eqn:rho}), (\ref{eqn:Qdef}), 
and (\ref{eqn:rdef}),
we obtain the means, variances, and covariance of $d$ and $y$ as follows:
\begin{align}
\lla d \rra 
&= \lla \sum_{i=1}^N g_i    u(n-i+1) \rra 
  = \sum_{i=1}^N g_i \lla u(n-i+1) \rra =0 , \label{eqn:mean_of_d} \\
\lla y \rra 
&= \lla \sum_{i=1}^N w_i u(n-i+1) \rra 
  = \sum_{i=1}^N w_i \lla u(n-i+1) \rra =0, \label{eqn:appy} \\
\lla d^2 \rra
&= \lla \lp \sum_{i=1}^N g_i    u(n-i+1) \rp^2 \rra 
  = \lla \sum_{i=1}^N \sum_{j=1}^N g_i g_j   u(n-i+1) u(n-j+1) \rra \nonumber \\
&= \sum_{i=1}^N g_i^2   \lla u(n-i+1)^2 \rra
  = \sigma^2 \sum_{i=1}^N g_i^2 
  \xrightarrow{N\rightarrow \infty} \rho^2 \sigma_g^2, \\  
\lla y^2 \rra
&= \lla \lp \sum_{i=1}^N w_i    u(n-i+1) \rp^2 \rra 
  = \lla \sum_{i=1}^N \sum_{j=1}^N w_i w_j   u(n-i+1) u(n-j+1) \rra \nonumber \\
&= \sum_{i=1}^N w_i^2   \lla u(n-i+1)^2 \rra
  = \sigma^2 \sum_{i=1}^N w_i^2 
  \xrightarrow{N\rightarrow \infty} \rho^2 Q, \label{eqn:appy2} \\  
\lla dy \rra
&= \lla \lp \sum_{i=1}^N g_i  u(n-i+1) \rp 
          \lp \sum_{j=1}^N w_j u(n-j+1) \rp \rra  
  = \lla \sum_{i=1}^N \sum_{j=1}^N g_i w_j   u(n-i+1) u(n-j+1) \rra \nonumber \\
&= \sum_{i=1}^N g_i w_i   \lla u(n-i+1)^2 \rra
  = \sigma^2 \sum_{i=1}^N g_i w_i 
  \xrightarrow{N\rightarrow \infty} \rho^2 r. \label{eqn:appdy}  
\end{align}

From (\ref{eqn:mean_of_d})--(\ref{eqn:appdy}),
the covariance matrix of $d$ and $y$ is (\ref{eqn:cov}).
Here, (\ref{eqn:appy}),  (\ref{eqn:appy2}), and  (\ref{eqn:appdy}) were derived
assuming that the correlation 
between $\bm{w}(n)$ and $\bm{u}(n)$
is small \cite{Costa2002,Tobias2000a,Tobias2000b}.  
This assumption is a standard assumption used to analyze
many adaptive algorithms\cite{Haykin2002,Sayed2003}.

\section{Derivation of (\ref{eqn:fy2})} \label{sec:appf(y)^2}
\begin{align}
\lla f(y)^2 \rra
&= \int_{-\infty}^\infty \mathrm{d}y f(y)^2 p(y) 
= \lp \int_{-\infty}^{-S} + \int_{-S}^{S}+ \int_{S}^{\infty} \rp
   \mathrm{d}y f(y)^2 
   \frac{1}{\sqrt{2\pi \rho^2 Q}}\exp \lp -\frac{y^2}{2\rho^2 Q}\rp 
   \nonumber \\
&= 2 
   \lp
   \underbrace{\int_{S}^{\infty} \mathrm{d}y S^2 
   \frac{1}{\sqrt{2\pi \rho^2 Q}}
   \exp \lp -\frac{y^2}{2\rho^2 Q}\rp}_{\ref{sec:appf(y)^2}1}
   + 
   \underbrace{\int_0^{S} \mathrm{d}y y^2 
   \frac{1}{\sqrt{2\pi \rho^2 Q}}
   \exp \lp -\frac{y^2}{2\rho^2 Q}\rp}_{\ref{sec:appf(y)^2}2}
   \rp,
\end{align}

\begin{align}
\ref{sec:appf(y)^2}1 
&= \int_{S}^{\infty} \mathrm{d}y S^2 
   \frac{1}{\sqrt{2\pi \rho^2 Q}}\exp \lp -\frac{y^2}{2\rho^2 Q}\rp 
=\frac{S^2}{\sqrt{\pi}} \int_{\frac{S}{\sqrt{2\rho^2 Q}}}^{\infty} 
  \mathrm{d}y'  \exp \lp -y'^2 \rp, 
	& \mbox{where } y'=\frac{y}{\sqrt{2\rho^2 Q}} \nonumber \\
&=\frac{S^2}{\sqrt{\pi}} 
   \lp \frac{\sqrt{\pi}}{2} -\int_0^{\frac{S}{\sqrt{2\rho^2 Q}}} \mathrm{d}y  \exp \lp -y^2 \rp \rp
   & \lp \because \int_0^{\infty} \mathrm{d}y  \exp \lp -y^2 \rp = \frac{\sqrt{\pi}}{2} \rp  \nonumber \\
&=\frac{S^2}{2}\lp 1-\mbox{erf}\lp \frac{S}{\sqrt{2\rho^2 Q}} \rp \rp,
\end{align}

\begin{align}
\ref{sec:appf(y)^2}2 
&= \int_{0}^{S} \mathrm{d}yy^2 
   \frac{1}{\sqrt{2\pi \rho^2 Q}}\exp \lp -\frac{y^2}{2\rho^2 Q}\rp \\
&=\left[- y \sqrt{\frac{\rho^2 Q}{2\pi}}\exp \lp -\frac{y^2}{2\rho^2 Q} \rp\right]_0^S
   + \sqrt{\frac{\rho^2 Q}{2\pi}} \int_{0}^{S} \mathrm{d}y 
   \exp \lp -\frac{y^2}{2\rho^2 Q}\rp, 
	& \mbox{where we used integration by parts } \\
&=-S\sqrt{\frac{\rho^2 Q}{2\pi}}\exp \lp -\frac{S^2}{2\rho^2 Q} \rp
   + \frac{\rho^2 Q}{2} \int_{0}^{\frac{S}{\sqrt{2\rho^2 Q}}} \mathrm{d}y' 
   \exp \lp -y'^2\rp, 
	& \mbox{where } y'=\frac{y}{\sqrt{2\rho^2 Q}} \nonumber \\
&=-S\sqrt{\frac{\rho^2 Q}{2\pi}}\exp \lp -\frac{S^2}{2\rho^2 Q} \rp
   + \frac{\rho^2 Q}{2} \mbox{erf}\lp \frac{S}{\sqrt{2\rho^2 Q}} \rp,  \label{eqn:J}
\end{align}

\begin{align}
\therefore 
\lla f(y)^2 \rra
&= 2 \lp \ref{sec:appf(y)^2}1+\ref{sec:appf(y)^2}2 \rp 
  =S^2-S\sqrt{\frac{\rho^2 Q}{2\pi}}\exp \lp -\frac{S^2}{2\rho^2 Q} \rp
   + \lp \rho^2 Q-S^2\rp \mbox{erf}\lp \frac{S}{\sqrt{2\rho^2 Q}} \rp.
\end{align}

\section{Derivation of (\ref{eqn:dfy})} \label{sec:appdf(y)}

\begin{align}
\lla df(y) \rra
&= \int_{-\infty}^\infty \int_{-\infty}^\infty 
   \mathrm{d}d \mathrm{d}y df(y) p(d,y) \\
&= \underbrace{\int_{-\infty}^\infty \mathrm{d}d d 
   \int_{-\infty}^{-S} \mathrm{d}y (-S) p(d,y)}_{\ref{sec:appdf(y)}1}
   +\underbrace{\int_{-\infty}^\infty \mathrm{d}d d \int_{-S}^S
    \mathrm{d}y yp(d,y)}_{\ref{sec:appdf(y)}2}
   +\underbrace{\int_{-\infty}^\infty \mathrm{d}d d \int_S^\infty
    \mathrm{d}y S p(d,y)}_{\ref{sec:appdf(y)}3},
\end{align}

\begin{align}
\ref{sec:appdf(y)}2
&= \int_{-\infty}^\infty \mathrm{d}d d \int_{-S}^S
    \mathrm{d}y y 
    \frac{1}{2\pi \sqrt{\left| \rho^2 \begin{pmatrix} \sigma_g^2 & r \\ r & Q \end{pmatrix} \right|}}
    \exp \lp -\frac{\begin{pmatrix} d & y \end{pmatrix}
    \lp \rho^2 \begin{pmatrix} \sigma_g^2 & r \\ r & Q \end{pmatrix}\rp^{-1}
    \begin{pmatrix} d \\ y \end{pmatrix}}
    {2}\rp \\
&= \int_{-\infty}^\infty \mathrm{d}d d \int_{-S}^S
    \mathrm{d}y y \frac{1}{2\pi \rho^2 \sqrt{\sigma_g^2 Q -r^2}}
    \exp \lp -\frac{Qd^2-2rdy+\sigma_g^2y^2}
    {2\rho^2 \lp \sigma_g^2 Q-r^2\rp}\rp \\
&= \int_{-\infty}^\infty \mathrm{d}d d \int_{-S}^S
    \mathrm{d}y y \frac{1}{2\pi \rho^2 \sqrt{\sigma_g^2 Q -r^2}}
    \exp \lp -\frac{Q\lp d-\frac{r}{Q}y\rp^2
    +\lp \sigma_g^2-\frac{r^2}{Q}\rp y^2}
    {2\rho^2 \lp \sigma_g^2 Q-r^2\rp}\rp \\
&= \int_{-S}^S \mathrm{d}y y 
    \exp \lp -\frac{y^2}{2\rho^2 Q}\rp
    \int_{-\infty}^\infty \mathrm{d}d d 
    \frac{1}{2\pi \rho^2 \sqrt{\sigma_g^2 Q -r^2}}
    \exp \lp -\frac{\lp d-\frac{r}{Q}y\rp^2}
    {2\rho^2 \lp \sigma_g^2-\frac{r^2}{Q}\rp}\rp \\
&=\int_{-S}^S \mathrm{d}y y 
    \exp \lp -\frac{y^2}{2\rho^2 Q}\rp
    \int_{-\infty}^\infty 
    \sqrt{2\rho^2 \lp \sigma_g^2-\frac{r^2}{Q}\rp}
    \mathrm{d}d'\lp \sqrt{2\rho^2 \lp \sigma_g^2-\frac{r^2}{Q}\rp}d'
    +\frac{r}{Q}y\rp \frac{1}{2\pi \rho^2 \sqrt{\sigma_g^2 Q -r^2}}
    \exp\lp -d'^2\rp, \\
	&\hspace{100mm} \mbox{where } d'=\frac{d-\frac{r}{Q}y}
	{\sqrt{2\rho^2 \lp \sigma_g^2-\frac{r^2}{Q}\rp}}\\
&=\int_{-S}^S \mathrm{d}y y 
    \exp \lp -\frac{y^2}{2\rho^2 Q}\rp
    \int_{-\infty}^\infty 
    2\rho^2 \lp \sigma_g^2-\frac{r^2}{Q}\rp
    \mathrm{d}dd\frac{1}{2\pi \rho^2 \sqrt{\sigma_g^2 Q -r^2}}
    \exp\lp -d^2\rp \nonumber \\
&\hspace{5mm}  +\int_{-S}^S \frac{r}{Q} \mathrm{d}y y^2 
    \exp \lp -\frac{y^2}{2\rho^2 Q}\rp
    \int_{-\infty}^\infty 
    \sqrt{2\rho^2 \lp \sigma_g^2-\frac{r^2}{Q}\rp}
    \mathrm{d}d
    \frac{1}{2\pi \rho^2 \sqrt{\sigma_g^2 Q -r^2}}
    \exp\lp -d^2\rp \\
&=\int_{-S}^S \frac{r}{Q} \mathrm{d}y y^2 
    \exp \lp -\frac{y^2}{2\rho^2 Q}\rp
    \int_{-\infty}^\infty 
    \mathrm{d}d
    \frac{1}{\pi \sqrt{2\rho^2 Q}}
    \exp\lp -d^2\rp
    \hspace{15mm} \lp \because \int_{-S}^S \mathrm{d}y y 
    \exp \lp -\frac{y^2}{2\rho^2 Q}\rp=0 \rp \nonumber
\end{align}
\begin{align}
&=\frac{2r}{Q\sqrt{2\pi \rho^2 Q}}
    \int_0^S \mathrm{d}y y^2 
    \exp \lp -\frac{y^2}{2\rho^2 Q}\rp
    \hspace{43mm} \lp \because \int_{-\infty}^\infty \mathrm{d}d \exp\lp -d^2\rp 
    =\sqrt{\pi} \rp \nonumber \\
&=\frac{2r}{Q\sqrt{2\pi \rho^2 Q}}
    \lp \left[ -\rho^2 Q y \exp \lp -\frac{y^2}{2\rho^2 Q}\rp \right]_0^S
    -
    \int_0^S \mathrm{d}y
    \lp -\rho^2 Q \rp 
    \exp \lp -\frac{y^2}{2\rho^2 Q}\rp \rp, 
	\\
	& \hspace{100mm} \mbox{where we used integration by parts } \nonumber \\
&=\frac{2r}{Q\sqrt{2\pi \rho^2 Q}}
    \lp -\rho^2 SQ \exp \lp -\frac{S^2}{2\rho^2 Q}\rp 
    +
    \rho^2 Q
    \int_0^\frac{S}{\sqrt{2\rho^2 Q}} 
    \sqrt{2\rho^2 Q}
    \mathrm{d}y'
    \exp \lp -y'^2 \rp \rp, 
     	\ \ \ \mbox{where }y'=\frac{y}{\sqrt{2\rho^2 Q}} \nonumber \\
&=-rS\rho \sqrt{\frac{2}{\pi Q}}
    \exp \lp -\frac{S^2}{2\rho^2 Q}\rp 
    +
    \rho^2 r\  
    \mbox{erf}\lp \frac{S}{\sqrt{2\rho^2 Q}} \rp,
\end{align}

\begin{align}
\ref{sec:appdf(y)}3
&= \int_{-\infty}^\infty \mathrm{d}d d \int_S^\infty
    \mathrm{d}y S 
    \frac{1}{2\pi \sqrt{\left| \rho^2 \begin{pmatrix} \sigma_g^2 & r \\ r & Q \end{pmatrix} \right|}}
    \exp \lp -\frac{\begin{pmatrix} d & y \end{pmatrix}
    \lp \rho^2 \begin{pmatrix} \sigma_g^2 & r \\ r & Q \end{pmatrix}\rp^{-1}
    \begin{pmatrix} d \\ y \end{pmatrix}}
    {2}\rp \\
&= \int_{-\infty}^\infty \mathrm{d}d d \int_S^\infty
    \mathrm{d}y S \frac{1}{2\pi \rho^2 \sqrt{\sigma_g^2 Q -r^2}}
    \exp \lp -\frac{Qd^2-2rdy+\sigma_g^2y^2}
    {2\rho^2 \lp \sigma_g^2 Q-r^2\rp}\rp \\
&= \int_{-\infty}^\infty \mathrm{d}d d \int_S^\infty
    \mathrm{d}y S \frac{1}{2\pi \rho^2 \sqrt{\sigma_g^2 Q -r^2}}
    \exp \lp -\frac{Q\lp d-\frac{r}{Q}y\rp^2
    +\lp \sigma_g^2-\frac{r^2}{Q}\rp y^2}
    {2\rho^2 \lp \sigma_g^2 Q-r^2\rp}\rp \\
&= S \int_S^\infty \mathrm{d}y  
    \exp \lp -\frac{y^2}{2\rho^2 Q}\rp
    \int_{-\infty}^\infty \mathrm{d}d d 
    \frac{1}{2\pi \rho^2 \sqrt{\sigma_g^2 Q -r^2}}
    \exp \lp -\frac{\lp d-\frac{r}{Q}y\rp^2}
    {2\rho^2 \lp \sigma_g^2-\frac{r^2}{Q}\rp}\rp \\
&=S \int_S^\infty \mathrm{d}y  
    \exp \lp -\frac{y^2}{2\rho^2 Q}\rp
    \int_{-\infty}^\infty 
    \sqrt{2\rho^2 \lp \sigma_g^2-\frac{r^2}{Q}\rp}
    \mathrm{d}d'\lp \sqrt{2\rho^2 \lp \sigma_g^2-\frac{r^2}{Q}\rp}d'+\frac{r}{Q}y\rp
    \frac{1}{2\pi \rho^2 \sqrt{\sigma_g^2 Q -r^2}}
    \exp\lp -d'^2\rp, \\
&  \hspace{100mm} \mbox{where } d'=\frac{d-\frac{r}{Q}y}
    {\sqrt{2\rho^2 \lp \sigma_g^2-\frac{r^2}{Q}\rp}}\\
%
&=S \int_S^\infty \mathrm{d}y 
    \exp \lp -\frac{y^2}{2\rho^2 Q}\rp
    \int_{-\infty}^\infty 
    2\rho^2 \lp \sigma_g^2-\frac{r^2}{Q}\rp
    \mathrm{d}dd\frac{1}{2\pi \rho^2 \sqrt{\sigma_g^2 Q -r^2}}
    \exp\lp -d^2\rp \nonumber \\
&\hspace{5mm}  +S\int_S^\infty \frac{r}{Q} \mathrm{d}y y 
    \exp \lp -\frac{y^2}{2\rho^2 Q}\rp
    \int_{-\infty}^\infty 
    \sqrt{2\rho^2 \lp \sigma_g^2-\frac{r^2}{Q}\rp}
    \mathrm{d}d
    \frac{1}{2\pi \rho^2 \sqrt{\sigma_g^2 Q -r^2}}
    \exp\lp -d^2\rp\\
&=\frac{Sr}{Q}\int_S^\infty \mathrm{d}y y 
    \exp \lp -\frac{y^2}{2\rho^2 Q}\rp
    \sqrt{2\rho^2 \lp \sigma_g^2-\frac{r^2}{Q}\rp}
    \frac{1}{2\pi \rho^2 \sqrt{\sigma_g^2 Q -r^2}}
    \sqrt{\pi} \\ 
    & \hspace{70mm} \lp \because 
    \int_{-\infty}^\infty \mathrm{d}dd \exp\lp -d^2\rp=0,\ \ 
    \int_{-\infty}^\infty \mathrm{d}d \exp\lp -d^2\rp 
    =\sqrt{\pi} \rp \\
&=\frac{Sr}{Q}\frac{1}{\sqrt{2\pi \rho^2 Q}}
    \left[ \lp -\rho^2 Q\rp \exp \lp -\frac{y^2}{2\rho^2 Q}\rp \right]_S^\infty
=\frac{Sr\rho}{\sqrt{2\pi Q}}\exp \lp -\frac{S^2}{2\rho^2 Q}\rp,
\end{align}

\begin{align}
\ref{sec:appdf(y)}1
&=\int_{-\infty}^\infty \mathrm{d}d d \int_{-\infty}^{-S} 
  \mathrm{d}y (-S) p(d,y)
=\ref{sec:appdf(y)}3, 
\hspace{10mm} \mbox{where we used integration by substitution:}\ y'=-y,\ d'=d, 
\end{align}

\begin{align}
\therefore
\lla df(y) \rra
&= \ref{sec:appdf(y)}1+\ref{sec:appdf(y)}2+\ref{sec:appdf(y)}3
  =\rho^2 r\  
    \mbox{erf}\lp \frac{S}{\sqrt{2\rho^2 Q}} \rp.
\end{align}

\section{Derivation of (\ref{eqn:yfy})} \label{sec:appyf(y)}

\begin{align}
\lla y(f(y))\rra
&=\int_{-\infty}^\infty \mathrm{d}y y f(y) p(y)
  = \underbrace{\int_{-\infty}^{-S} \mathrm{d}y y (-S) p(y)}_{\ref{sec:appyf(y)}1}
   +\underbrace{\int_{-S}^{S} \mathrm{d}y y^2 p(y)}_{\ref{sec:appyf(y)}2}
   +\underbrace{\int_S^\infty \mathrm{d}y y S p(y)}_{\ref{sec:appyf(y)}3},
\end{align}

\begin{align}
\ref{sec:appyf(y)}2
&=-S\sqrt{\frac{2\rho^2 Q}{\pi}}\exp \lp -\frac{S^2}{2\rho^2 Q} \rp
    + \rho^2 Q\ \mbox{erf}\lp \frac{S}{\sqrt{2\rho^2 Q}} \rp,  
    & \lp \because (\ref{eqn:J})\rp
\end{align}

\begin{align}
\ref{sec:appyf(y)}1&=\ref{sec:appyf(y)}3
=S \int_S^\infty \mathrm{d}y y \frac{1}{\sqrt{2\pi \rho^2 Q}}
    \exp \lp -\frac{y^2}{2\rho^2 Q} \rp 
=\frac{S}{\sqrt{2\rho^2 Q}}\left[-\rho^2 Q 
   \exp \lp -\frac{y^2}{2\rho^2 Q} \rp\right]_S^\infty
=S\sqrt{\frac{\rho^2 Q}{2\pi}}\exp \lp -\frac{S^2}{2\rho^2 Q} \rp.
\end{align}

\begin{align}
\therefore
\lla yf(y)\rra
&=\ref{sec:appyf(y)}1+\ref{sec:appyf(y)}2+\ref{sec:appyf(y)}3 
  =\rho^2 Q\ \mbox{erf}\lp \frac{S}{\sqrt{2\rho^2 Q}}\rp.
\end{align}

\section{Derivation of (\ref{eqn:R=1})} \label{sec:appR=1}
When $S<S_C$,
since $Q\rightarrow \infty$ in the limit $t\rightarrow \infty$,
from (\ref{eqn:drdt2}), (\ref{eqn:Qr}), and (\ref{eqn:erf_approx}),
we obtain
\begin{align}
\frac{\mathrm{d}r}{\mathrm{d}t}
&=\sigma_g \cos \theta \frac{d\sqrt{Q}}{\mathrm{d}t} 
=\mu \rho^2 
\left(\sigma_g^2-\sigma_g \cos \theta \sqrt{Q}\ 
   \mbox{erf}\left(\frac{S}{\sqrt{2\rho^2 Q}} \right)\right)
\simeq \mu \rho^2 
\left(\sigma_g^2-\sigma_g \cos \theta \sqrt{Q}\ 
   \frac{2}{\sqrt{\pi}} \frac{S}{\sqrt{2\rho^2 Q}} \right),
\label{eqn:drdt3}\\
\therefore \frac{d\sqrt{Q}}{\mathrm{d}t}
&\simeq \mu \rho^2 
\left(\frac{\sigma_g}{\cos \theta}-\sqrt{\frac{2}{\pi\rho^2}}S \right).
\label{eqn:dldt}
\end{align}

On the other hand, from 
(\ref{eqn:dQdt2}), (\ref{eqn:Qr}), (\ref{eqn:erf_approx}), and (\ref{eqn:exp_approx}),
we obtain
\begin{align}
\frac{\mathrm{d}Q}{\mathrm{d}t}
&=\frac{\mathrm{d}\lp\sqrt{Q}\rp^2}{\mathrm{d}t}
  =2\sqrt{Q}\frac{d\sqrt{Q}}{\mathrm{d}t}\\
&\simeq \mu \rho^2 
\Bigl(\mu \left(\rho^2 Q-2\rho^2 \sigma_g \sqrt{Q} \cos \theta-S^2\right)-2Q\Bigr)
\frac{2}{\sqrt{\pi}} \frac{S}{\sqrt{2\rho^2 Q}}
- \mu^2 \rho^2 S \sqrt{\frac{2\rho^2 Q}{\pi}}\left(1-\frac{S^2}{2\rho^2 Q}\right)
\nonumber \\
&\hspace{4mm} 
+\mu \rho^2 \left(\mu\left(\rho^2 \sigma_g^2 + S^2+\sigma_\xi^2\right)
   +2\sigma_g \sqrt{Q} \cos \theta\right).
\label{eqn:dQdt3}
\end{align}
The right-hand side of the formula, 
which is obtained by dividing both sides of (\ref{eqn:dQdt3}) by $2\sqrt{Q}$,
should be equal to the right-hand side of (\ref{eqn:dldt}).
Therefore, solving the equation for $\cos \theta$,
we obtain 
\begin{align}
\cos \theta &\xrightarrow{t\rightarrow \infty}1.
\end{align}

\section{Derivation of (\ref{eqn:MSE3})} \label{sec:appMSE3}
When $S<S_C$,
$Q\rightarrow \infty$ and $\cos \theta =1$ in the limit $t\rightarrow \infty$.
Therefore, from (\ref{eqn:dldt}), we obtain
\begin{align}
\frac{\mathrm{d}\sqrt{Q}}{\mathrm{d}t}
&\simeq \mu \rho^2 
\left(\sigma_g-\sqrt{\frac{2}{\pi\rho^2}}S \right), \label{eqn:dldt2} \\
\therefore \sqrt{Q}
&\simeq \mu \rho^2 
\left(\sigma_g-\sqrt{\frac{2}{\pi\rho^2}}S \right)t+\mbox{Const.}
\label{eqn:l}
\end{align}
Note that (\ref{eqn:l}) represents the divergence of $\sqrt{Q}$
with time, which does not contradict 
the fact that 
no solutions were found for the simultaneous equations that are obtained by 
substituting zeros into the left-hand sides
of the simultaneous differential equations (\ref{eqn:drdt2}) and (\ref{eqn:dQdt2})
at $S<S_C$ in Sec. \ref{sec:steady_state}.

On the other hand, from (\ref{eqn:Qr}), we obtain
\begin{align}
r&\simeq \sigma_g \sqrt{Q}. \label{eqn:Qr2}
\end{align}
By substituting (\ref{eqn:erf_approx}), (\ref{eqn:exp_approx}), 
(\ref{eqn:l}), and (\ref{eqn:Qr2}) into (\ref{eqn:MSE2})
and 
arranging the expression,
we obtain the MSE as 
\begin{align}
\lla e^2 \rra
&=
S^2-2\sigma_g \rho \sqrt{\frac{2}{\pi}} S 
+ \sigma_g^2 \rho^2 + \sigma_\xi^2.
\label{eqn:MSE4}
\end{align}

\twocolumn

\begin{IEEEbiography}{Seiji Miyoshi}
received his B.Eng. and M.Eng. degrees in
electrical engineering from Kyoto University, Japan, in 1986 and 1988,
respectively, and his Ph.D degree in system science and engineering
from Kanazawa University, Japan, in 1998. He worked  with the
Space Development Division of
NEC Corporation from 1988 to 1994. He joined the Department of
Electronic Engineering of Kobe City College of Technology in 1994.
He joined
the Department of Electrical and Electronic Engineering,
Faculty of Engineering Science of Kansai University in 2008,
where he is now a professor. His research interests
include statistical-mechanical informatics, image processing,
signal processing, and learning theory.
He is
a senior member of
the Institute of Electronics, Information and Communication Engineers (IEICE)
and
the Institute of Electrical and Electronics Engineers (IEEE),
and
a member of
the Physical Society of Japan (JPS)
and
the Japan Neural Network Society (JNNS).
He served as an associate editor of
IEICE Transactions on Fundamentals of Electronics,
Communications and Computer Sciences
from 2013 to 2017
and an awards committee member of
the IEEE Kansai section from 2011 to 2014.
He received
the 2011 Papers of Editors' Choice from the Journal of the Physical
Society of Japan,
the Excellent Paper Award in ICONIP2016,
and
the 2019 Best Paper Award from IEICE.
\end{IEEEbiography}



\end{document}